\documentclass[symmetry,article,accept,pdftex,moreauthors]{Definitions/mdpi} 

\firstpage{1} 
\pubvolume{1}
\issuenum{1}
\articlenumber{0}
\pubyear{2025}
\copyrightyear{2025}
\externaleditor{Firstname Lastname}
\datereceived{14 December 2024} 
\daterevised{20 January 2025} % Comment out if no revised date
\dateaccepted{22 January 2025} 
\datepublished{ } 
\hreflink{https://doi.org/} % If needed use \linebreak
\usepackage{natbib,fontenc}
\newcommand{\aap}{Astron. Astrophys.}
\newcommand{\aj}{Astron. J.}
\newcommand{\apj}{Astrophys. J.}
\newcommand{\apjs}{Astrophys. J. Suppl. Ser.}

\newcommand{\mnras}{Mon. Not. R. Astron. Soc.}
\newcommand{\aapr}{Astron. Astrophys. Rev.}
\newcommand{\aaps}{Astron. Astrophys. Suppl.}
\newcommand{\pasp}{Publ. Astron. Soc. Pac.}

\newcommand{\araa}{Annu. Rev. Astron. Astrophys.}
\newcommand{\apss}{Astrophys. Space Sci.}

\Title{What is Inside the Double--Double Structure of the Radio Galaxy J0028+0035?}

\TitleCitation{What is Inside the Double--Double Structure of the Radio Galaxy J0028+0035?}

\Author{S\'andor Frey $^{1,2,3,}$*\orcidA{}, Andrzej Marecki $^{4}$\orcidB{}, Krisztina \'Eva Gab\'anyi $^{1,2,5,6}$\orcidC{} and Marek Jamrozy $^{7}$\orcidD{}}

\AuthorNames{S\'andor Frey, Andrzej Marecki, Krisztina \'Eva Gab\'anyi and Marek Jamrozy}

\AuthorCitation{Frey, S.; Marecki, A.; Gab\'anyi, K.\'E.; Jamrozy, M.}
\address{%
$^{1}$ \quad Konkoly Observatory, HUN-REN Research Centre for Astronomy and Earth Sciences, Konkoly Thege \mbox{Mikl\'os \'ut 15-17}, 1121 Budapest, Hungary; k.gabanyi@astro.elte.hu\\

$^{2}$ \quad CSFK, MTA Centre of Excellence, Konkoly Thege Mikl\'os \'ut 15-17, H-1121 Budapest, Hungary\\
$^{3}$ \quad Institute of Physics and Astronomy, ELTE E\"otv\"os Lor\'and University, P\'azm\'any P\'eter s\'et\'any 1/A, \mbox{1117 Budapest, Hungary}\\
$^{4}$ \quad Institute of Astronomy, Faculty of Physics, Astronomy and Informatics,  Nicolaus Copernicus University, \mbox{ul. Grudzi\k{a}dzka 5,}
87-100 Toru\'n, Poland; amr@astro.umk.pl\\
$^{5}$ \quad Department of Astronomy, Institute of Physics and Astronomy, ELTE E\"otv\"os Lor\'and University, \mbox{P\'azm\'any P\'eter s\'et\'any 1/A,} 1117 Budapest, Hungary\\
$^{6}$ \quad HUN-REN---ELTE Extragalactic Astrophysics Research Group, ELTE E\"otv\"os Lor\'and University, \mbox{P\'azm\'any P\'eter s\'et\'any 1/A,} 1117 Budapest, Hungary\\
$^{7}$ \quad Astronomical Observatory, Jagiellonian University, Orla 171, 30-244 Krak\'ow, Poland; jamrozy@oa.uj.edu.pl}

\corres{Correspondence: frey.sandor@csfk.org}

\abstract{The radio source J0028+0035 is a recently discovered double--double radio galaxy at redshift $z=0.398$. Its relic outer lobes are separated by about $3^{\prime}$ in the sky, corresponding to $\sim$1~Mpc projected linear size. Inside this large-scale structure, the inner pair of collinear lobes span about $100$~kpc. In the arcsec-resolution radio images of J0028+0035, there is a central radio feature that offers the intriguing possibility of being resolved into a pc-scale, third pair of innermost lobes. This would make this radio galaxy a rare triple--double source where traces of three distinct episodes of radio activity could be observed. To reveal the compact radio structure of the central component, we conducted observation with the European Very Long Baseline Interferometer Network and the enhanced Multi Element Remotely Linked Interferometer Network. Our $1.66$ GHz image with high (\mbox{$\sim$5 milliarcsec}) resolution shows a compact central radio core with no indication of a third, innermost double feature. The observation performed in multi-phase-centre mode also revealed that the physically unrelated but in projection closely separated background source 5BZU\,J0028+0035 has a single weak, somewhat resolved radio feature, at odds with its blazar classification.}


\begin{document}

%%%%%%%%%%%%%%%%%%%%%%%%%%%%%%%%%%%%%%%%%%
\section{Introduction}
\label{sec:intro}

Owing to their powerful emission across the entire electromagnetic spectrum, active galactic nuclei (AGNs) are distant signposts in the Universe, e.g., \cite{2017A&ARv..25....2P}. They are powered by accretion onto supermassive black holes, {BHs}, located in the centres of galaxies. Nearly one-tenth of AGNs, e.g., \cite{2002AJ....124.2364I}, have powerful bipolar plasma jets where the charged particles spiralling around the magnetic field lines with relativistic speeds produce synchrotron radiation. It makes these objects especially luminous in the radio waveband. They are called radio-loud or jetted \cite{2017NatAs...1E.194P} AGNs. 
The appearance and observational properties of radio-loud AGNs strongly depend on the orientation of their jets with respect to our line of \mbox{sight \cite{1995PASP..107..803U}}. In blazar-type AGNs, one side of the symmetric pair of jets emanating from the vicinity of the {BH} is closely ($\lesssim$$10^{\circ}$) aligned with the viewing direction, leading to Doppler-boosted emission. On the contrary, radio galaxy jets lie close to the plane of the sky. 

Highly collimated jets in radio galaxies can travel significantly beyond the visible boundaries of their host galaxies, extending up to several Mpc in size. Giant radio galaxies are conventionally defined as being larger than $0.7$~Mpc \cite{2023JApA...44...13D}. However, the jets do not extend indefinitely, as they interact with the surrounding intergalactic medium, eventually transferring sufficient energy and momentum to the ambient medium to substantially reduce their propagation speed. Here, the plasma carried by the jets all along from the central AGN spreads out and forms extended lobes on both sides of the galaxy. The lobes are filled with relativistic plasma and are magnetized, continuing to emit synchrotron radiation. The continuous supply for these reservoirs of magnetic fields and high-energy charged particles can stop when the nuclear activity terminates and thus jets are no longer fed. Consequently, the radio emission of the lobes gradually fades out.  In the synchrotron process, higher-energy electrons lose energy faster, thus the radio spectrum of the source steepens over time at higher frequencies. As a consequence of spectral aging, the initial power-law shape develops a spectral break and steepens over time, e.g., \cite{1991ApJ...383..554C}. This way, the past history of central activity is imprinted in the structural and spectral properties of \mbox{the lobes}.

Returning to the central engine, intermittent jet activity can be caused by alternating phases of accretion and quiescence of the supermassive {BH} powering the AGN. During active phases, powerful relativistic jets are produced, while in quiescent phases, jet production ceases, though previously ejected plasma still radiates as it cools. There is ample evidence for the intermittent nature of the AGN phenomenon. Activity periods can last for a total of $\sim$$10^{7-9}$~years during the whole lifetime of a galaxy, but are most likely decomposed into a series of much shorter periods that can be as short as $\sim$$10^{5}$~years \cite{2015MNRAS.451.2517S}. Radio galaxy lobe structures may demonstrate in the most definite way two or even three successive episodes of nuclear activity, in the form of double--double or triple--double sources. In these cases, two or three pairs of distinct radio lobes are observed. The aging synchrotron-radiating plasma in the outermost lobes created during an earlier activity episode is still observable as diffuse steep-spectrum radio emission, while the renewed nuclear activity produces the inner pair(s) of lobes closer to the galactic centre. If the recurrent activity starts before the total disappearance of the outer lobe emission, we can observe nested, collinear pairs of radio lobes. In double--double radio galaxies (DDRGs), there is evidence for two activity periods, e.g., \cite{2000MNRAS.315..371S,2009BASI...37...63S,2011MNRAS.410..484B,2012BASI...40..121N,2017MNRAS.471.3806K,2019A&A...622A..13M,2021MNRAS.501..853M}. There are more than a hundred DDRGs known to date. The discovery of the faint, diffuse relic lobes requires sensitive imaging at low radio frequencies. Triple--double sources with evidence of two previous and one current activity episodes are much more rare; only four of them are known at present \cite{2007MNRAS.382.1019B,2011MNRAS.417L..36H,2016ApJ...826..132S,2023MNRAS.525L..87C}.

\begin{figure}[H]
%\centering
\includegraphics[width=11 cm]{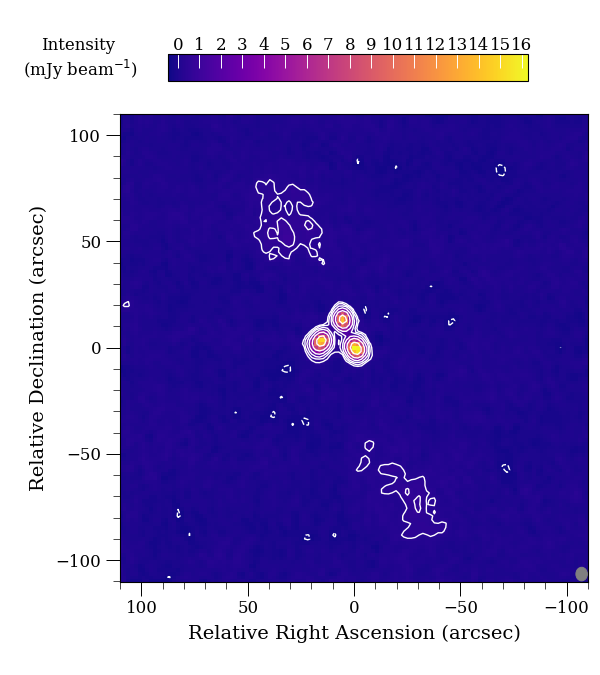}
\caption{The $1.4$ GHz FIRST \cite{1995ApJ...450..559B} image of the DDRG J0028+0035, together with the Roma-BZCAT source 5BZU\,J0028+0035 (the easternmost one in the central trio of compact sources on the arcsec scale). The symmetric pair of relic outer lobes is seen as faint extended features towards the northeast and southwest. The peak intensity is $16.3$~mJy\,beam$^{-1}$. The lowest contours are drawn at $\pm0.34$~mJy\,beam$^{-1}$ ($\sim$$3\sigma$ image noise). The positive contour levels increase by a factor of $2$. The half-power width of the elliptical Gaussian restoring beam is $5.4^{\prime\prime} \times 6.4^{\prime\prime}$ at major axis position angle $\mathrm{PA} = 0^{\circ}$, as indicated with an ellipse in the lower-right corner. \label{fig:first}} 
\end{figure}

The double--double radio galaxy J0028+0035 was discovered by \cite{2021MNRAS.501..853M} when investigating the large-scale radio structure of the blazar 5BZU\,J0028+0035 from the Roma-BZCAT multifrequency catalogue \cite{2009A&A...495..691M,2015Ap&SS.357...75M}. The corresponding optical source is known as SDSS\,J002839.77+003542.2 \cite{2008ApJS..175..297A} or SDSS\,J002839.77+003542.1 \cite{2017ApJS..233...25A}. The $1.4$ GHz radio image from the Faint Images of the Radio Sky at Twenty-cm (FIRST) survey \cite{1995ApJ...450..559B,1997ApJ...475..479W} indicated two symmetric relic lobes straddling the source that was accompanied by two other, similarly compact components within $\sim$$20^{\prime\prime}$ (Figure~\ref{fig:first}). A thorough analysis of archival radio data as well as new higher-resolution radio images revealed that the two additional compact components are in fact the inner lobes of a Fanaroff--Riley (FR) type II \cite{1974MNRAS.167P..31F} radio galaxy. The relic outer lobes also belong to this object, making J0028+0035 a double--double radio galaxy \cite{2021MNRAS.501..853M}. It is identified with the optical galaxy SDSS\,J002838.86+003539.7 \cite{2008ApJS..175..297A}, also known as SDSS\,J002838.85+003539.8 at a spectroscopic redshift of $z_\mathrm{G}=0.398$ \cite{2017ApJS..233...25A}. This is one of the possible candidates for the brightest galaxy {(BCG)} of the cluster number 6035 in the optical catalogue of \cite{2011ApJ...736...21S}. On the other hand, the source 5BZU\,J0028+0035 appears as a background AGN seen coincidentally close in projection to the radio galaxy. Its redshift is $z_\mathrm{B}=0.686$ \cite{2017ApJS..233...25A}, excluding the possibility of a physical relationship with the double--double radio galaxy.

According to \cite{2021MNRAS.501..853M}, the observed properties of J0028+0035 resemble a typical DDRG. The estimated ages of the outer and inner lobes are $2.45 \times 10^{8}$~years and $3.6 \times 10^{6}$~years, \mbox{respectively}. \textls[-15]{The time length of quiescence between the two active phases was \mbox{$1.1 \times 10^{7}$~years}. (These age estimates are believed to be accurate to $\sim$$20\%$.)
Our general understanding of the physical processes that trigger restarted jet activity is still limited \cite{2023Galax..11...74M}. According to observations, AGN jets are disrupted on a broad range of timescales, from $\sim$$10^4$~years to $\sim$$10^7$~years. This suggests that various mechanisms may be at work that influence jet (re)triggering. These include instabilities of the accretion flow and accretion rate variability, variations of the magnetic field of the accretion disk, in some cases galaxy interactions and mergers that can funnel gas towards the central supermassive {BH}s, and {BH} spin changes either caused suddenly by mergers or gradually by prolonged accretion episodes. }

{The spin plays a crucial role in jet production since the rotational energy of the supermassive BH can be extracted to power relativistic jets through the Blandford--Znajek mechanism} \cite{1977MNRAS.179..433B}. {BH spins can change with time as a consequence of accretion, as well as mergers during the hierarchical formation of galaxies and their central BH} \cite{2005ApJ...620...69V}. {However, in the case of the DDRG J0028+0035, there is no evidence for a recent BH merger, especially because the position angles of the outer and inner lobes, at least as projected onto the sky, are similar} \cite{2021MNRAS.501..853M}. {Since jets are expelled along the rotation axis, it is unlikely that a spin flip caused by a major merger}, e.g., \cite{2009ApJ...697.1621G}, {occurred when the jet activity restarted. Therefore, possible effects of accretion are considered only in the following. Coherent accretion (i.e., when the angular momentum of the inflowing gas aligns with that of the BH) can spin up the BH to the theoretical maximum (i.e., the dimensionless spin parameter $a$ almost reaching unity) within $\sim$$10^8$~years. On the other hand, chaotic (misaligned) accretion in a gas-poor environment can reduce BH spin on a timescale of $\sim$$10^8$--$10^9$~years, depending on the actual accretion rate. In comparison with the ages estimated by} \cite{2021MNRAS.501..853M} for the radio features in DDRG J0028+0035, it is feasible that the activity cycle responsible for the relic outer lobes ceased because of BH spin-down due to chaotic accretion at a sub-Eddington rate. On the other hand, the central BH could regain enough momentum via coherent accretion during its radio-quiet period of $\sim$$10^7$~years that is sufficient for jet production.
Powerful jets are known to require high BH spin ($a \gtrsim 0.5$). However, moderate values ($a \gtrsim 0.1$) can already be enough for producing relativistic jets in magnetically arrested accretion \mbox{flows \cite{2011MNRAS.418L..79T}}. {While the timescales are broadly consistent with possible changes in the BH spin, this effect is not necessarily the main driver behind the subsequent periods of jet activity and quiescence in the DDRG J0028+0035, especially because, after all, the spin may not be a dominant parameter for powering jets. Instead, state changes in the accretion disk may be responsible for jet production} \cite{2013A&A...557L...7V}. {By simulating the evolution of massive BH spins on cosmological timescales,} ref. \cite{2005ApJ...620...69V} {also found that high spin is not a necessary and sufficient condition for jetted AGNs. In a study of jet activity in BCGs,} ref. \cite{2011ApJ...727...39M} were not able to tell whether gas accretion or BH spin was the dominant factor in powering jets, but their data are consistent with a broad range of spin parameters and accretion rates. It is more likely that intrinsic effects like magnetic flux variations or changes in the mass accretion rate within the accretion system cause the jets to switch off and on with a similar \mbox{power \cite{2019A&A...622A..13M}.}

In the double--double radio sources, the inner lobes evolve into an almost uniform medium inside a cocoon surrounding them, formed by the jet flow during the previous active phase of the nucleus that produced the outer lobes. The largest linear size between the outer lobes exceeds $1$~Mpc, making J0028+0035 a giant radio galaxy \cite{2020A&A...642A.153D,2023JApA...44...13D}. The inner lobes span nearly $100$~kpc. The asymmetries seen in the structure may be caused by jet reorientation between the subsequent activity episodes or/and the inhomogeneity of the surrounding intergalactic medium.

\begin{figure}[H]
%\centering
\includegraphics[width=11 cm]{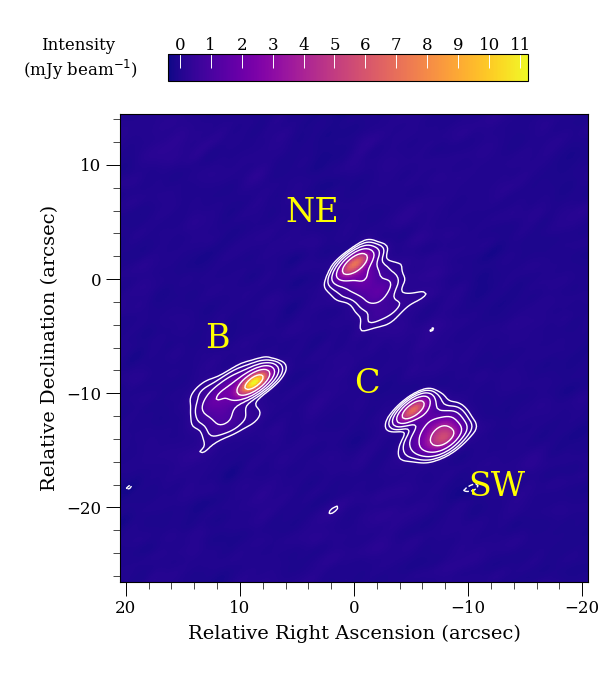}
\caption{The $1.52$ GHz VLA A-configuration image of the Roma-BZCAT source 5BZU\,J0028+0035 (labeled as B) and the DDRG J0028+0035, whose core, northeastern, and southwestern inner lobes are marked with C, NE, and SW, respectively. The image is adopted from Figure~5 in \cite{2021MNRAS.501..853M}. The peak intensity is $11.3$~mJy\,beam$^{-1}$. The lowest contours are drawn at $\pm0.25$~mJy\,beam$^{-1}$. The positive contour levels increase by a factor of $2$. The half-power width of the elliptical Gaussian restoring beam is $1.0^{\prime\prime} \times 2.5^{\prime\prime}$ at $\mathrm{PA} = -52^{\circ}$. The coordinates are relative to the phase centre at right ascension $00^{\mathrm{h}}\,28^{\mathrm{min}}\,39.164^{\mathrm{s}}$ and declination $+00^{\circ}\,35^{\prime}\,51.61^{\prime\prime}$. Note that this reference position is different from the one in Figure~\ref{fig:first}.} \label{fig:vla-a} 
\end{figure}

The highest-resolution images of \cite{2021MNRAS.501..853M} made at $1.5$~GHz (restoring beam size: $1.0^{\prime\prime} \times 2.5^{\prime\prime}$, see also in Figure~\ref{fig:vla-a}) and $5.5$~GHz ($0.3^{\prime\prime} \times 0.75^{\prime\prime}$) with the Karl G. Jansky Very Large Array (VLA) indicate that the southwestern inner lobe-like feature is clearly separated into an actual faint diffuse lobe and a bright compact central component that remains unresolved on these scales. The latter core component positionally coincides with the optical galaxy, suggesting that the AGN itself is located here.     

To reveal the nature of the arcsec-scale compact core in the DDRG J0028+0035, we performed radio imaging observation with even higher angular resolution, $\sim$5 milliarcseconds (mas), using the technique of very long baseline interferometry (VLBI) involving $17$ globally distributed radio telescopes of the European VLBI Network (EVN) and the British enhanced Multi Element Remotely Linked Interferometer Network (e-MERLIN) at $1.66$~GHz frequency. We also observed the nearby Roma-BZCAT source 5BZU\,J0028+0035 seen in projection within a small angular distance from the DDRG core. In Section~\ref{sec:obs}, we describe the observation and the data reduction. The results are presented in Section~\ref{sec:res} and discussed in Section~\ref{sec:disc}. Finally, we summarise our findings in Section~\ref{sec:concl}. 

In this paper, to calculate linear sizes and luminosity distances, we assume the same flat $\Lambda$ Cold Dark Matter cosmological model as \cite{2021MNRAS.501..853M}, with Hubble constant \mbox{$H_0=71$~km\,s$^{-1}$\,Mpc$^{-1}$,} vacuum energy density parameter $\Omega_\Lambda = 0.73$, and matter density parameter $\Omega_\mathrm{m}= 0.27$. In this model, the angular scale at the redshift $z_\mathrm{G}=0.398$ of the DDRG J0028+0035 is $5.327$~kpc\,arcsec$^{-1}$ \cite{2006PASP..118.1711W}.

%%%%%%%%%%%%%%%%%%%%%%%%%%%%%%%%%%%%%%%%%%
\section{VLBI Observation and Data Reduction}
\label{sec:obs}

\subsection{Phase-Referenced EVN Observation in Multi-Phase-Centre Mode}

The observation with the EVN and e-MERLIN (project code: EM152, PI: A. Marecki) took place on 21 May 2021 at the central frequency of $1.66$~GHz. The experiment lasted for $4$~h. The bandwidth was divided into four adjacent $32$ MHz wide intermediate frequency channels (IFs), each consisting of 64 spectral channels. Recording was conducted in both left and right circular polarizations, with a $1024$~Mbps data rate. 

The observation was scheduled in phase-referencing mode \citep{1995ASPC...82..327B}. This involved regular nodding between a nearby calibrator source and the weak target, to allow for coherent integration on the latter well exceeding the atmospheric coherence time. Phase referencing also provided accurate position determination for the target relative to the calibrator source. We used $580$ s long phase-reference cycles, with $460$~s spent on the target sources. The total on-target time was $2.8$~h. We selected the compact quasar J0029$-$0113 ($1.82^{\circ}$ separation) as the phase-reference calibrator. Its accurate astrometric position (right ascension $\alpha_\mathrm{ref} = 00^{\mathrm{h}}\,29^{\mathrm{min}}\, (00.98603 \pm 0.00001)^{\mathrm{s}}$, declination $\delta_\mathrm{ref} =  -01^{\circ}\,13^{\prime}\,(41.7597 \pm 0.0003)^{\prime\prime}$; reference epoch J2015.0) is listed in the third edition of the Internationial Celestial Reference \mbox{Frame \cite{2020A&A...644A.159C}} catalogue (\url{https://hpiers.obspm.fr/icrs-pc/newwww/icrf/icrf3sx.txt}, accessed on \mbox{14 December 2024}). The bright radio source 3C\,454.3 was also occasionally observed as a fringe-finder. 

Our primary target was the core component of the DDRG J0028+0035 (referred to as C hereafter; see Figure~\ref{fig:vla-a} as a finding chart), revealed earlier in the arcsec-resolution VLA A-configuration images \cite{2021MNRAS.501..853M}. Given that the nearly coincident Roma-BZCAT source 5BZU\,J0028+0035 (referred to as B) is located within only $\sim$$15^{\prime\prime}$ in projection, we utilized the multi-phase-centre observing mode \cite{2011PASP..123..275D,2014A&A...563A.111C,2015ExA....39..259K}. Since the sources fell well within the primary beam of even the largest EVN antennas \cite{2014A&A...563A.111C}, the same antenna pointings could be used for observing them. At the correlation, separate passes with different a priori positions were made. We were also granted correlation at a third phase centre, positioned at the brighter northeastern (NE) inner lobe of J0028+0035 \cite{2021MNRAS.501..853M}, to see whether any mas-scale compact radio emission from a putative hotspot could be detected there. This way, the EVN observation and correlation allowed us to investigate three different very nearby targets with the same antenna pointings. The data were processed at the Joint Institute for VLBI European Research Infrastructure Consortium (JIVE, Dwingeloo, The Netherlands) with the SFXC software correlator \cite{2015ExA....39..259K} using $2$~s integration time.

The following 17 radio telescopes participated in the experiment and provided useful data (their antenna diameters are given in parentheses): Jodrell Bank Mk2 ($38\,\mathrm{m} \times 25\,\mathrm{m}$, United Kingdom), Westerbork ($25$~m, The Netherlands), Effelsberg ($100$~m, Germany), Medicina ($32$~m, Italy), Onsala ($25$~m, Sweden), Tianma ($65$~m, China), Toru\'{n} ($32$~m, Poland), Hartebeesthoek ($26$~m, South Africa), Svetloe ($32$~m, Russia), Zelenchukskaya ($32$~m, Russia), Badary ($32$~m, Russia), Irbene ($32$~m, Latvia), Sardinia ($65$~m, Italy), and the e-MERLIN antennas in the United Kingdom: Cambridge ($32$~m), Darnhall, Knockin, and Pickmere ($25$~m each). 

\subsection{Data Calibration and Imaging}

The raw correlated visibility data were loaded into the U.S. National Radio Astronomy Observatory (NRAO) Astronomical Image Processing System (\textsc{aips}) \cite{2003ASSL..285..109G} and calibrated following the standard steps, e.g., \cite{1995ASPC...82..227D}. The amplitudes were calibrated using the antenna gain curves and system temperatures ($T_\mathrm{sys}$) measured at the telescopes. For the e-MERLIN stations, nominal $T_\mathrm{sys}$ values had to be used. Data points obtained when antennas were off-source were flagged. The dispersive ionospheric delays were determined using total electron content maps derived from global navigation satellite systems data. Phase corrections due to source parallactic angle variations (for azimuth--elevation mounted antennas), and manual instrumental phase and delay corrections using a short $1$ min scan on the phase-reference calibrator source J0029$-$0113 were performed. As a simple bandpass correction, the first and last four frequency channels of each IF were flagged because of low amplitudes and noisier phases. After performing global fringe-fitting \citep{1983AJ.....88..688S} on the strong phase-reference (J0029$-$0113) and fringe-finder (3C\,454.3) sources, the calibrated visibility data were exported to the Caltech \textsc{Difmap} software package \citep{1994BAAS...26..987S} for hybrid mapping. This involved several iterations of \textsc{clean} deconvolution \citep{1974A&AS...15..417H} and phase-only \mbox{self-calibration \citep{1984ARA&A..22...97P},} followed by iterations of amplitude and phase self-calibration with solution intervals gradually decreasing from the whole observing period to zero. The antenna-based gain correction factors determined using the two strong sources were averaged and, after returning to \textsc{aips}, applied to the visibility amplitudes if they exceeded the absolute value of $5\%$. Fringe-fitting was then repeated for the phase-reference calibrator J0029$-$0113, now taking its \textsc{clean} component model determined in \textsc{Difmap} into account, to correct for potential source structure contribution to the phases. The final fringe-fit solutions obtained for the calibrator were interpolated to the target source data (C, B, and NE) for all three corresponding correlator passes separately. 

The calibrated visibility data of the target sources were transferred to  \textsc{Difmap} for imaging and brightness distribution modelling. Two target sources (C and B) were detected as they produced a signal in the dirty images. Their visibility data were modelled using a single circular Gaussian brightness distribution component \citep{1995ASPC...82..267P} in \textsc{Difmap}. The model parameters were adjusted during the hybrid mapping process involving phase-only self-calibration steps, with gradually decreasing solution intervals from $30$ to $2$~min. Only selected sensitive European antennas (Effelsberg, Sardinia, Onsala, Jodrell Bank, and Toru\'n) were set to correctable; phases at the other antennas were held fixed. Uncertainties of the final Gaussian model parameter were estimated following \cite{1999ASPC..180..301F,2008AJ....136..159L}. For the flux densities, an additional $5\%$ uncertainty was added in quadrature because of the VLBI absolute amplitude calibration uncertainty. 

The phase-referenced positions of the targets detected were determined by locating their brightness peaks with the \textsc{aips} command \textsc{maxfit}. The estimated astrometric accuracy is $\sim$1~mas in both right ascension and declination.

%%%%%%%%%%%%%%%%%%%%%%%%%%%%%%%%%%%%%%%%%%
\section{Results}
\label{sec:res}

\subsection{The Core of the Radio Galaxy J0028+0035}

\begin{figure}[H]
\includegraphics[width=9 cm]{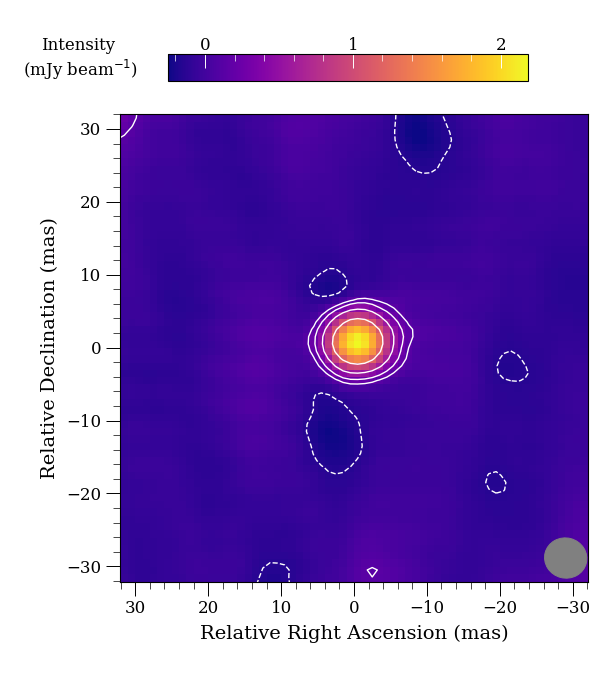}
\caption{The naturally weighted $1.66$ GHz EVN image of the DDRG J0028+0035 core (C). The peak intensity is $2.18$~mJy\,beam$^{-1}$. The lowest contours are drawn at $\pm0.14$~mJy\,beam$^{-1}$ ($\sim$$3\sigma$ image noise). The positive contour levels increase by a factor of $2$. The half-power width of the elliptical Gaussian restoring beam is $5.5\,\mathrm{mas} \times 5.8\,\mathrm{mas}$ at $\mathrm{PA} = 76^{\circ}$. \label{fig:ddrg-evn}} 
\end{figure}

Our primary target, the core of the DDRG J0028+0035 (C), was detected at right ascension $\alpha_\mathrm{C,VLBI} = 00^{\mathrm{h}}\,28^{\mathrm{min}}\,38.85220^{\mathrm{s}}$ and declination $\delta_\mathrm{C,VLBI} = +00^{\circ}\,35^{\prime}\,39.7843^{\prime\prime}$. Its VLBI image shows a single component (Figure~\ref{fig:ddrg-evn}). The parameters of the circular Gaussian brightness distribution model, the $1.66$ GHz flux density ($S$), and the full width at half maximum (FWHM) diameter ($\theta$), are listed in Table~\ref{tab:model}. We derived the brightness temperature, \mbox{e.g., \cite{1982ApJ...252..102C}},
\begin{equation}
T_{\mathrm{b}} = 1.22 \times 10^{12} \, (1 + z) \frac{S}{\theta^2 \nu^2} \,\, \mathrm{[K]},
\label{eq:tb}
\end{equation}
and the monochromatic radio power
\begin{equation}
       P_{\nu} = 4\pi D_{\mathrm{L}}^2 S (1+z)^{-1-\alpha}.
\label{eq:pw}
\end{equation}
In Equation~(\ref{eq:tb}) above, $z$ is the redshift. The observing frequency $\nu$ is measured in GHz (\mbox{in our case,} $\nu=1.66$~GHz), $S$ in Jy, and $\theta$ in mas. In Equation~\ref{eq:pw}, $D_{\mathrm{L}}$ denotes the luminosity distance. Here, the parameters are expressed in SI units, [$P_{\nu}$]=W\,Hz$^{-1}$, [$D_{\mathrm{L}}$]=m, and \mbox{[$S$]= W\,m$^{-2}$\,Hz$^{-1}$}. In the absence of VLBI measurements at any other frequency, for the radio power-law spectral index $\alpha$ (according to the definition $S \propto \nu^{\alpha}$), we assume the approximate value of $-0.9$ inferred from the multi-frequency flux density measurements collected by \cite{2021MNRAS.501..853M} on the arcsec scale. (Note that the exact value of $\alpha$ has moderate influence on the power calculation because the source is at relatively low redshift, $z_\mathrm{G}=0.398$.)

\subsection{The Northeastern Inner Lobe of J0028+0035}

In our high-resolution VLBI imaging observation, we did not detect the northeastern inner radio galaxy lobe (NE). The brightness upper limit derived from the dirty image was $0.26$~mJy\,beam$^{-1}$, corresponding to $6\sigma$ noise level. Assuming a coherence loss of \mbox{$\sim$$25\%$ \cite{2010A&A...515A..53M,2019A&A...630L...5G,2022ApJS..260...49K,2024A&A...690A.321K},} we can conclude that there was no $\lesssim10$-mas-scale compact component (hotspot) embedded in this lobe above $0.33$~mJy flux density. In other words, the extended emission with a total $\sim$18~mJy flux density measured at this frequency band on arcsec \mbox{scale \cite{2021MNRAS.501..853M}} is completely resolved out on the EVN and e-MERLIN baselines.

\subsection{The Radio AGN 5BZU\,J0028+0035}

\begin{figure}[H]
%\centering
\includegraphics[width=9 cm]{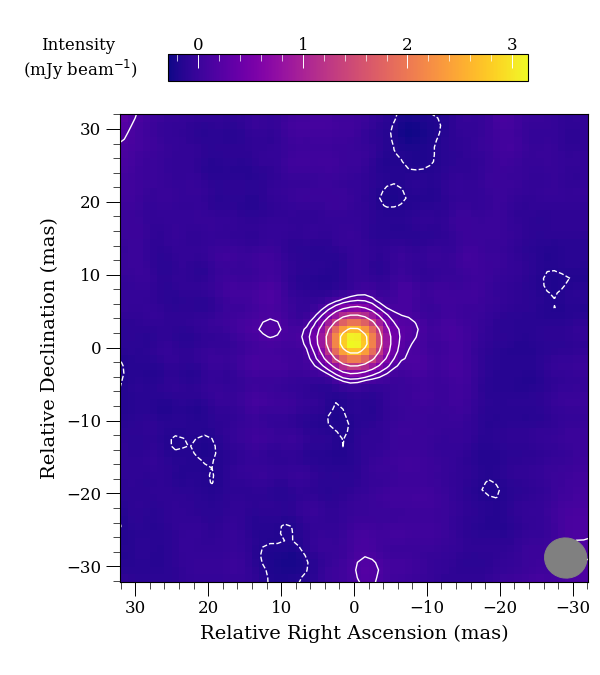}
\caption{The naturally weighted $1.66$ GHz EVN image of 5BZU\,J0028+0035 (B). The peak intensity is $3.15$~mJy\,beam$^{-1}$. The lowest contours are drawn at $\pm0.16$~mJy\,beam$^{-1}$ ($\sim$$3\sigma$ image noise). The positive contour levels increase by a factor of $2$. The half-power width of the elliptical Gaussian restoring beam is $5.5\,\mathrm{mas} \times 5.8\,\mathrm{mas}$ at $\mathrm{PA} = 77^{\circ}$. \label{fig:5bzu-evn}} 
\end{figure}

Our secondary target in chance coincidence with the radio galaxy, the Roma-BZCAT source 5BZU\,J0028+0035 (B), was also detected with the EVN. Its right ascension and declination were $\alpha_\mathrm{B,VLBI} = 00^{\mathrm{h}}\,28^{\mathrm{min}}\,39.77257^{\mathrm{s}}$ and $\delta_\mathrm{B,VLBI} = +00^{\circ}\,35^{\prime}\,42.2164^{\prime\prime}$, respectively. Similarly to the radio galaxy core, we detected a single mas-scale compact component (Figure~\ref{fig:5bzu-evn}). The parameters of the circular Gaussian brightness distribution model and the derived physical parameters are given in Table~\ref{tab:model}. To calculate the rest-frame \mbox{$1.66$ GHz} monochromatic power, we assumed the power-law spectral index $-0.6$ based on multi-frequency total flux density measurements \cite{2021MNRAS.501..853M}.

\begin{table}[H] 
\caption{Parameters of the circular Gaussian model components describing the brightness distribution of the DDRG J0028+0035 core (C, top) and the Roma-BZCAT source 5BZU\,J0028+0035 (B, bottom), along with their derived physical parameters.\label{tab:model}}
\tablesize{\fontsize{9}{9}\selectfont}
%\newcolumntype{C}{>{\centering\arraybackslash}X}
\begin{tabularx}{\textwidth}{m{1cm}<{\centering}CCm{4cm}<{\centering}C}
\toprule
\textbf{Source}	& \textbf{Flux Density}	& \textbf{Size FWHM} & \textbf{Brightness Temperature} & \textbf{Radio Power}\\
            	& \textbf{(mJy)}	    & \textbf{(mas)} & \textbf{(K)} & \textbf{(W\,Hz$^{-1}$)}\\

\midrule
C		& $3.7\pm0.6$			& $4.0\pm0.6$			& $(1.4 \pm 1.0) \times 10^{8} $			& $(2.0 \pm 0.3) \times 10^{24}$ \\
B		& $5.1\pm0.9$			& $3.5\pm0.5$			& $(3.1\pm2.1) \times 10^{8} $			& $(8.6 \pm 1.5) \times 10^{24}$ \\
\bottomrule
\end{tabularx}
%\noindent{\footnotesize{\textsuperscript{1} Tables may have a footer.}}
\end{table}

%%%%%%%%%%%%%%%%%%%%%%%%%%%%%%%%%%%%%%%%%%
\section{Discussion}
\label{sec:disc}

\subsection{The Radio Galaxy J0028+0035}

The brightness temperature of the single component detected at mas scale ($T_\mathrm{b} \sim 10^8$~K, Table~\ref{tab:model}) clearly indicates that its radio emission is of AGN origin since it cannot be reconciled with any plausible extended emission from the host galaxy \cite{2000ApJ...530..704K}. The monochromatic radio power ($P_\mathrm{1.66\,GHz} \sim 10^{24}$~W\,Hz$^{-1}$) is typical for low-luminosity radio AGN, e.g., \cite{2020Symm...12..527K,2023MNRAS.518...39W}. Concerning its position, an optical counterpart of the DDRG J0028+0035 core is not found in the \textit{Gaia} space astrometry mission \cite{2016A&A...595A...1G} Data Release 3 (DR3) \cite{2023A&A...674A...1G} catalogue. However, the coordinates of this weak optical source ($r$ magnitude $19.08\pm0.04$) in the Sloan Digital Sky Survey (SDSS) DR13 \cite{2017ApJS..233...25A} ($\alpha_\mathrm{C,SDSS} = 00^{\mathrm{h}}\,28^{\mathrm{min}}\,38.8555^{\mathrm{s}}, \delta_\mathrm{C,SDSS} = +00^{\circ}\,35^{\prime}\,39.803^{\prime\prime}$; \url{https://skyserver.sdss.org/dr13/en/tools/explore/Summary.aspx?id=1237663716017439748}, \mbox{accessed on 14 December 2024}) are consistent with our VLBI position, given that the SDSS astrometry is known to be accurate to $\sim$$50$~mas in both \mbox{coordinates \cite{2013A&A...553A..13O}.} Thus, we did not find an innermost double-lobe structure that would have served as a clear evidence for a distinct third, most recent activity cycle, nor a mas-scale jet pointing towards the arcsec-scale lobes. The single weak central component may indicate that the source is currently in transition from radio-loud to radio-quiet state \cite{2011A&A...525A...6M}. 

The host of the radio galaxy J0028+0035 is also detected by the \textit{Wide-field Infrared Survey Explorer} (\textit{WISE}) satellite \cite{2010AJ....140.1868W} in the mid-infrared. Its magnitudes in the $W1$ and $W2$ bands, as well as magnitude limits in the $W3$ and $W4$ bands as given in the AllWISE catalogue (\cite{2014yCat.2328....0C}, \url{https://cdsarc.cds.unistra.fr/viz-bin/cat/II/328}, accessed on 14 December 2024), are listed in Table~\ref{tab:wise}.

According to the VLA imaging \cite{2021MNRAS.501..853M}, the two inner lobes (NE and SW) show highly asymmetric properties. While NE seems to be an FR-II type lobe, SW has a faint, diffuse appearance. However, according to our VLBI observation, there is no evidence for a compact hotspot in NE either. The different morphologies of the two inner lobes may naturally be explained if they are no longer actively fueled by a jet. In this case, their apparent asymmetry results from light-travel time lag: because of the inclination of the system with respect to the plane of the sky, the NE lobe is farther away from us than the SW one, so we observe it at an earlier evolutionary stage when its radio emission is less decayed than that of the SW lobe. For a detailed explanation of this phenomenon, see \cite{2012A&A...544L...2M}.

The difference in the core flux density between the arcsec-scale VLA ($8.4\pm0.9$~mJy at $1.52$~GHz, \cite{2021MNRAS.501..853M}) and our mas-scale VLBI measurements ($3.7 \pm 0.6$~mJy at 1.66~GHz, Table~\ref{tab:model}) can be attributed to the presence of diffuse emission on $\sim$0.1--1$^{\prime\prime}$ angular scale, resolved out on the long interferometer baselines. Furthermore, there may be some coherence loss in our VLBI imaging \cite{2010A&A...515A..53M} since self-calibration could not be performed for all telescopes.  

\begin{table}[H] 
\caption{\textit{WISE} mid-infrared magnitudes of the DDRG J0028+0035 (top) and the Roma-BZCAT source 5BZU\,J0028+0035 (bottom).\label{tab:wise}}
%\newcolumntype{C}{>{\centering\arraybackslash}X}
\begin{tabularx}{\textwidth}{CCCCC}
\toprule
\textbf{Source}	& \boldmath{$W1$} \textbf{Band} & \textbf{ \boldmath{$W2$} Band} & \textbf{ \boldmath{$W3$} Band} & \textbf{ \boldmath{$W4$} Band}\\
            	& \boldmath{$3.35$~$\upmu$\textbf{m}}      & \boldmath{$4.6$~$\upmu$\textbf{m}}        & \boldmath{$11.6$~$\upmu$\textbf{m}}      & \boldmath{$22.1$~$\upmu$\textbf{m}}  \\

\midrule
DDRG J0028+0035  & $14.93\pm0.04$	 & $14.79\pm0.07$  	  & $>11.76$ 		   & $>8.94$ \\
5BZU\,J0028+0035 & $14.31\pm0.03$	 & $14.06\pm0.05$     & $>11.99$		   & $>8.42$ \\
\bottomrule
\end{tabularx}
%\noindent{\footnotesize{\textsuperscript{1} Tables may have a footer.}}
\end{table}

\subsection{The Blazar Candidate 5BZU\,J0028+0035}

The putative blazar 5BZU\,J0028+0035 also shows a single mas-scale component in our $1.66$ GHz VLBI image (Figure~\ref{fig:5bzu-evn}). Both the brightness temperature and the monochromatic power values are in the same order of magnitude as measured for the DDRG\,J0028+0035 core (Table~\ref{tab:model}). Blazars are known to have relativistic jets pointing nearly to the line of sight, resulting in Doppler-boosted radio emission. For blazars, the Doppler factor, i.e., the ratio of the measured and the intrinsic brightness temperatures ($\Delta = T_\mathrm{b}/T_\mathrm{b,int}$), exceeds unity. Since $T_\mathrm{b,int}$$\sim$$10^{10}$~K \cite{1994ApJ...426...51R,2021ApJ...923...67H}, and the measured brightness temperature for 5BZU\,J0028+0035 is two orders of magnitude smaller than that, we found $\Delta \ll 1$ and thus no evidence for the blazar nature of this source. 

The definition of blazars is quite vague and can be more liberal in other wavebands (see a recent discussion in \cite{2024Galax..12....8K}). For the Roma-BZCAT catalogue \cite{2009A&A...495..691M}, the set of selection criteria includes radio detection down to mJy-level flux densities, compact radio structure (albeit generally on $\sim$1--10$^{\prime\prime}$ scales only), the availability of optical spectrum, and X-ray detection. The letter ``U'' in the designation of 5BZU\,J0028+0035 indicates an uncertain/transitional type. It is known that, under radio scrutiny, not all Roma-BZCAT sources qualify as blazars, even among the high-confidence ones \cite{2024ApJ...964...98X}. 

As evidenced by the VLA A-configuration observations, especially at $1.5$~GHz \mbox{(Figure~\ref{fig:vla-a}) \cite{2021MNRAS.501..853M},} the arcsec-scale structure of 5BZU\,J0028+0035 shows a jet-like extension towards east. By comparing our $1.66$ GHz flux density ($5.1 \pm 0.9$~mJy, Table~\ref{tab:model}) with the VLA value ($21.3 \pm 2.2$~mJy, \cite{2021MNRAS.501..853M}), it is most likely that a large fraction of the total flux density comes from extended structures resolved out on VLBI baselines. Alternatively, high-amplitude variability, a characteristic property of blazars, might also explain the flux density difference between the VLA and EVN measurements. However, there is no indication of variability from  multi-epoch flux-density data \cite{2021MNRAS.501..853M}. Finally, the steep radio spectrum ($\alpha \approx -0.6$, \cite{2021MNRAS.501..853M}) is also atypical for blazars.

It is worth mentioning that, similarly to the radio galaxy host, 5BZU\,J0028+0035 also has mid-infrared detection by \textit{WISE} (Table~\ref{tab:wise}). These values place the source outside of the AGN wedge defined on the $(W1-W2)$ vs. $(W2-W3)$ colour--colour diagram \cite{2012MNRAS.426.3271M}, and also below the $\gamma$-ray blazar strip \cite{2012ApJ...750..138M}. We consider these facts as additional indications against the blazar nature of 5BZU\,J0028+0035. 

The most accurate optical position of this object from the \textit{Gaia} space astrometry mission \cite{2016A&A...595A...1G} DR3 \cite{2023A&A...674A...1G} is right ascension $\alpha_\mathrm{B,Gaia} = 00^{\mathrm{h}}\,28^{\mathrm{min}}\,39.77278^{\mathrm{s}}$ and declination $\delta_\mathrm{B,Gaia} =  +00^{\circ}\,35^{\prime}\,42.2173^{\prime\prime}$ (reference epoch J2016.0), with formal errors of $0.45$ and $0.34$~mas, respectively. However, the non-zero astrometric excess noise \cite{2012A&A...538A..78L} of $1.07$~mas suggests the presence of modelling errors that are not accounted for by the observational noise. Its significance value ($0.95$) is below the threshold of $2$, therefore the source may still be astrometrically reliable (see the \textit{Gaia} DR3 documentation at \url{https://gea.esac.esa.int/archive/documentation/GDR3/pdf/GaiaDR3_documentation_1.3.pdf}, accessed on 14 December 2024). The \textit{Gaia} and VLBI positions of 5BZU\,J0028+0035 differ by $\sim$3~mas in right ascension and $\sim$1~mas in declination, therefore agree within reasonable uncertainties. 

%%%%%%%%%%%%%%%%%%%%%%%%%%%%%%%%%%%%%%%%%%
\section{Summary and Conclusions}
\label{sec:concl}

We observed the core (C) and the northeastern (NE) inner lobe of the recently discovered double--double radio galaxy J0028+0035 \cite{2021MNRAS.501..853M} at $1.66$~GHz with the high-resolution VLBI technique using the EVN and e-MERLIN arrays. While mas-scale compact radio emission was not detected from the lobe, the galaxy core showed a single emission feature at a celestial position consistent with that of the optical host galaxy. The measured brightness temperature and monochromatic radio power are typical of a low-luminosity radio AGN. Together with the arcsec-scale radio data, the lack of mas-scale jet emission suggests that the DDRG\,J0028+0035 is in transition from radio-loud to radio-quiet state, and the inner lobes are no longer actively supplied by the central engine.

Multi-phase-centre correlation of VLBI data allowed us to image the Roma-BZCAT blazar candidate 5BZU\,J0028+0035 (B) as well. This background AGN is physically unrelated to the DDRG but located inside the primary beam of the radio telescopes. A single mas-scale radio component was detected in accurate positional coincidence with the corresponding optical AGN. Our measurements indicate low brightness temperature and low radio power. Together with other radio and mid-infrared evidence, our results suggest that 5BZU\,J0028+0035 is in fact not a blazar.

%%%%%%%%%%%%%%%%%%%%%%%%%%%%%%%%%%%%%%%%%%
\vspace{6pt} 

%%%%%%%%%%%%%%%%%%%%%%%%%%%%%%%%%%%%%%%%%%
\authorcontributions{Conceptualization, A.M., S.F., K.\'E.G., and M.J.; methodology, A.M., S.F., K.\'E.G., and M.J.; formal analysis, S.F.; writing---original draft preparation, S.F.; writing---review and editing, A.M., K.\'E.G., and M.J.; visualization, S.F. All authors have read and agreed to the published version of the manuscript.}

\funding{The research leading to these results has received funding from the European Union's Horizon 2020 Research and Innovation Programme under grant agreement No. 101004719 (OPTICON RadioNet Pilot). This research was funded by the Hungarian National Research, Development and Innovation Office (NKFIH, grant no. OTKA K134213), the NKFIH excellence grant TKP2021-NKTA-64, and the HUN-REN Hungarian Research Network.}

\dataavailability{The observational data are in the public domain and available in the EVN Data Archive (\url{https://archive.jive.nl/}, accessed on 14 December 2024). Calibrated data will be made available upon reasonable request to the corresponding author.} 

\acknowledgments{We are grateful for the comments received from the three anonymous referees. The EVN is a joint facility of independent European, African, Asian, and North American radio astronomy institutes. Scientific results from data presented in this publication are derived from the following EVN project code: EM152. The e-MERLIN is a National Facility operated by the University of Manchester at Jodrell Bank Observatory on behalf of STFC. 
This work presents results from the European Space Agency (ESA) space mission \textit{Gaia}. \textit{Gaia} data are being processed by the \textit{Gaia} Data Processing and Analysis Consortium (DPAC). Funding for the DPAC is provided by national institutions, in particular the institutions participating in the \textit{Gaia} MultiLateral Agreement (MLA). The \textit{Gaia} mission website is \url{https://www.cosmos.esa.int/gaia}, accessed on 14 December 2024. The \textit{Gaia} archive website is \url{https://archives.esac.esa.int/gaia}, accessed on 14 December 2024.

This publication made use of data products from the \textit{Wide-field Infrared Survey Explorer}, which is a joint project of the University of California, Los Angeles, and the Jet Propulsion Laboratory/California Institute of Technology, funded by the National Aeronautics and Space Administration. 
This research has made use of the NASA/IPAC Extragalactic Database (NED) which is operated by the Jet Propulsion Laboratory, California Institute of Technology, under contract with the National Aeronautics and Space Administration. 
We thank Krisztina Perger for assistance in generating the figures with radio images.}

\conflictsofinterest{The authors declare no conflicts of interest.} 

%%%%%%%%%%%%%%%%%%%%%%%%%%%%%%%%%%%%%%%%%%
%% Optional

\abbreviations{Abbreviations}{
The following abbreviations are used in this manuscript:\\

\noindent 
\begin{tabular}{@{}ll}
AGN & active galactic nuclei\\
AIPS & Astronomical Image Processing System\\
{BCG} & {brightest cluster galaxy}\\
{BH} & {black hole}\\
DDRG & double--double radio galaxy\\
DR & data release\\
e-MERLIN & enhanced Multi Element Remotely Linked Interferometer Network\\
EVN & European VLBI Network\\
FIRST & Faint Images of the Radio Sky at Twenty cm\\
FR & Fanaroff--Riley type\\ 
FWHM & full width at half maximum\\
IF & intermediate frequency channel\\
mas & milliarcsecond\\
NRAO & U.S. National Radio Astronomy Observatory\\
PA & position angle\\
SDSS & Sloan Digital Sky Survey\\
SI & International System of Units\\
VLA & Karl G. Jansky Very Large Array\\
VLBI & very long baseline interferometry
\end{tabular}
}

%%%%%%%%%%%%%%%%%%%%%%%%%%%%%%%%%%%%%%%%%%
\begin{adjustwidth}{-\extralength}{0cm}
%\printendnotes[custom] % Un-comment to print a list of endnotes

\reftitle{References}

% Please provide either the correct journal abbreviation (e.g., according to the “List of Title Word Abbreviations” http://www.issn.org/services/online-services/access-to-the-ltwa/) or the full name of the journal.
% Citations and References in Supplementary files are permitted provided that they also appear in the reference list here. 

%=====================================
% References, variant A: external bibliography
%=====================================
%\bibliography{double-double}

\begin{thebibliography}{999}

\bibitem[{Padovani} et~al.(2017){Padovani}, {Alexander}, {Assef}, {De Marco},
  {Giommi}, {Hickox}, {Richards}, {Smol{\v{c}}i{\'c}}, {Hatziminaoglou},
  {Mainieri}, and {Salvato}]{2017A&ARv..25....2P}
{Padovani}, P.; {Alexander}, D.M.; {Assef}, R.J.; {De Marco}, B.; {Giommi}, P.;
  {Hickox}, R.C.; {Richards}, G.T.; {Smol{\v{c}}i{\'c}}, V.; {Hatziminaoglou},
  E.; {Mainieri}, V.;  et~al.
\newblock {Active galactic nuclei: what is in a name?}
\newblock {\em \aapr} {\bf 2017}, {\em 25}, 2. 
\newblock {\url{https://doi.org/10.1007/s00159-017-0102-9}}.

\bibitem[{Ivezi{\'c}} et~al.(2002){Ivezi{\'c}}, {Menou}, {Knapp}, {Strauss},
  {Lupton}, {Vanden Berk}, {Richards}, {Tremonti}, {Weinstein}, {Anderson},
  {Bahcall}, {Becker}, {Bernardi}, {Blanton}, {Eisenstein}, {Fan},
  {Finkbeiner}, {Finlator}, {Frieman}, {Gunn}, {Hall}, {Kim}, {Kinkhabwala},
  {Narayanan}, {Rockosi}, {Schlegel}, {Schneider}, {Strateva}, {SubbaRao},
  {Thakar}, {Voges}, {White}, {Yanny}, {Brinkmann}, {Doi}, {Fukugita},
  {Hennessy}, {Munn}, {Nichol}, and {York}]{2002AJ....124.2364I}
{Ivezi{\'c}}, {\v{Z}}.; {Menou}, K.; {Knapp}, G.R.; {Strauss}, M.A.; {Lupton},
  R.H.; {Vanden Berk}, D.E.; {Richards}, G.T.; {Tremonti}, C.; {Weinstein},
  M.A.; {Anderson}, S.;  et~al.
\newblock {Optical and Radio Properties of Extragalactic Sources Observed by
  the FIRST Survey and the Sloan Digital Sky Survey}.
\newblock {\em \aj} {\bf 2002}, {\em 124},~2364--2400.
\newblock {\url{https://doi.org/10.1086/344069}}.

\bibitem[{Padovani}(2017)]{2017NatAs...1E.194P}
{Padovani}, P.
\newblock {On the two main classes of active galactic nuclei}.
\newblock {\em Nature Astronomy} {\bf 2017}, {\em 1},~0194.
\newblock {\url{https://doi.org/10.1038/s41550-017-0194}}.

\bibitem[{Urry} and {Padovani}(1995)]{1995PASP..107..803U}
{Urry}, C.M.; {Padovani}, P.
\newblock {Unified Schemes for Radio-Loud Active Galactic Nuclei}.
\newblock {\em \pasp} {\bf 1995}, {\em 107},~803.
\newblock {\url{https://doi.org/10.1086/133630}}.

\bibitem[{Dabhade} et~al.(2023){Dabhade}, {Saikia}, and
  {Mahato}]{2023JApA...44...13D}
{Dabhade}, P.; {Saikia}, D.J.; {Mahato}, M.
\newblock {Decoding the giant extragalactic radio sources}.
\newblock {\em J. Astrophys. Astron.} {\bf 2023}, {\em 44},~13.
\newblock {\url{https://doi.org/10.1007/s12036-022-09898-5}}.

\bibitem[{Carilli} et~al.(1991){Carilli}, {Perley}, {Dreher}, and
  {Leahy}]{1991ApJ...383..554C}
{Carilli}, C.L.; {Perley}, R.A.; {Dreher}, J.W.; {Leahy}, J.P.
\newblock {Multifrequency Radio Observations of Cygnus A: Spectral Aging in
  Powerful Radio Galaxies}.
\newblock {\em \apj} {\bf 1991}, {\em 383},~554.
\newblock {\url{https://doi.org/10.1086/170813}}.


\bibitem[{Schawinski} et~al.(2015){Schawinski}, {Koss}, {Berney}, and
  {Sartori}]{2015MNRAS.451.2517S}
{Schawinski}, K.; {Koss}, M.; {Berney}, S.; {Sartori}, L.F.
\newblock {Active galactic nuclei flicker: an observational estimate of the
  duration of black hole growth phases of {\ensuremath{\sim}}{}10$^{5}$ yr}.
\newblock {\em \mnras} {\bf 2015}, {\em 451},~2517--2523.
\newblock {\url{https://doi.org/10.1093/mnras/stv1136}}.

\bibitem[{Schoenmakers} et~al.(2000){Schoenmakers}, {de Bruyn},
  {R{\"o}ttgering}, {van der Laan}, and {Kaiser}]{2000MNRAS.315..371S}
{Schoenmakers}, A.P.; {de Bruyn}, A.G.; {R{\"o}ttgering}, H.J.A.; {van der
  Laan}, H.; {Kaiser}, C.R.
\newblock {Radio galaxies with a `double-double morphology' - I. Analysis of
  the radio properties and evidence for interrupted activity in active galactic
  nuclei}.
\newblock {\em \mnras} {\bf 2000}, {\em 315},~371--380.
\newblock {\url{https://doi.org/10.1046/j.1365-8711.2000.03430.x}}.

\bibitem[{Saikia} and {Jamrozy}(2009)]{2009BASI...37...63S}
{Saikia}, D.J.; {Jamrozy}, M.
\newblock {Recurrent activity in Active Galactic Nuclei}.
\newblock {\em Bull. Astron. Soc. India} {\bf 2009}, {\em
  37},~63--89.
\newblock {\url{https://doi.org/10.48550/arXiv.1002.1841}}.

\bibitem[{Brocksopp} et~al.(2011){Brocksopp}, {Kaiser}, {Schoenmakers}, and {de
  Bruyn}]{2011MNRAS.410..484B}
{Brocksopp}, C.; {Kaiser}, C.R.; {Schoenmakers}, A.P.; {de Bruyn}, A.G.
\newblock {Double-double radio galaxies: further insights into the formation of
  the radio structures}.
\newblock {\em \mnras} {\bf 2011}, {\em 410},~484--498.
\newblock {\url{https://doi.org/10.1111/j.1365-2966.2010.17456.x}}.

\bibitem[{Nandi} and {Saikia}(2012)]{2012BASI...40..121N}
{Nandi}, S.; {Saikia}, D.J.
\newblock {Double-double radio galaxies from the FIRST survey}.
\newblock {\em Bull. Astron. Soc. India} {\bf 2012}, {\em
  40},~121--137.
\newblock {\url{https://doi.org/10.48550/arXiv.1208.1941}}.

\bibitem[{Ku{\'z}micz} et~al.(2017){Ku{\'z}micz}, {Jamrozy},
  {Kozie{\l}-Wierzbowska}, and {We{\.z}gowiec}]{2017MNRAS.471.3806K}
{Ku{\'z}micz}, A.; {Jamrozy}, M.; {Kozie{\l}-Wierzbowska}, D.; {We{\.z}gowiec},
  M.
\newblock {Optical and radio properties of extragalactic radio sources with
  recurrent jet activity}.
\newblock {\em \mnras} {\bf 2017}, {\em 471},~3806--3826.
\newblock {\url{https://doi.org/10.1093/mnras/stx1830}}.

\bibitem[{Mahatma} et~al.(2019){Mahatma}, {Hardcastle}, {Williams}, {Best},
  {Croston}, {Duncan}, {Mingo}, {Morganti}, {Brienza}, {Cochrane},
  {G{\"u}rkan}, {Harwood}, {Jarvis}, {Jamrozy}, {Jurlin}, {Morabito},
  {R{\"o}ttgering}, {Sabater}, {Shimwell}, {Smith}, {Shulevski}, and
  {Tasse}]{2019A&A...622A..13M}
{Mahatma}, V.H.; {Hardcastle}, M.J.; {Williams}, W.L.; {Best}, P.N.; {Croston},
  J.H.; {Duncan}, K.; {Mingo}, B.; {Morganti}, R.; {Brienza}, M.; {Cochrane},
  R.K.;  et~al.
\newblock {LoTSS DR1: Double-double radio galaxies in the HETDEX field}.
\newblock {\em \aap} {\bf 2019}, {\em 622},~A13.
\newblock {\url{https://doi.org/10.1051/0004-6361/201833973}}.

\bibitem[{Marecki} et~al.(2021){Marecki}, {Jamrozy}, {Machalski}, and
  {Pajdosz-{\'S}mierciak}]{2021MNRAS.501..853M}
{Marecki}, A.; {Jamrozy}, M.; {Machalski}, J.; {Pajdosz-{\'S}mierciak}, U.
\newblock {Multifrequency study of a double-double radio galaxy J0028+0035}.
\newblock {\em \mnras} {\bf 2021}, {\em 501},~853--865.
\newblock {\url{https://doi.org/10.1093/mnras/staa3632}}.

\bibitem[{Brocksopp} et~al.(2007){Brocksopp}, {Kaiser}, {Schoenmakers}, and {de
  Bruyn}]{2007MNRAS.382.1019B}
{Brocksopp}, C.; {Kaiser}, C.R.; {Schoenmakers}, A.P.; {de Bruyn}, A.G.
\newblock {Three episodes of jet activity in the Fanaroff-Riley type II radio
  galaxy B0925+420}.
\newblock {\em \mnras} {\bf 2007}, {\em 382},~1019--1028.
\newblock {\url{https://doi.org/10.1111/j.1365-2966.2007.12483.x}}.

\bibitem[{Hota} et~al.(2011){Hota}, {Sirothia}, {Ohyama}, {Konar}, {Kim},
  {Rey}, {Saikia}, {Croston}, and {Matsushita}]{2011MNRAS.417L..36H}
{Hota}, A.; {Sirothia}, S.K.; {Ohyama}, Y.; {Konar}, C.; {Kim}, S.; {Rey},
  S.C.; {Saikia}, D.J.; {Croston}, J.H.; {Matsushita}, S.
\newblock {Discovery of a spiral-host episodic radio galaxy}.
\newblock {\em \mnras} {\bf 2011}, {\em 417},~L36--L40.
\newblock {\url{https://doi.org/10.1111/j.1745-3933.2011.01115.x}}.

\bibitem[{Singh} et~al.(2016){Singh}, {Ishwara-Chandra}, {Kharb}, {Srivastava},
  and {Janardhan}]{2016ApJ...826..132S}
{Singh}, V.; {Ishwara-Chandra}, C.H.; {Kharb}, P.; {Srivastava}, S.;
  {Janardhan}, P.
\newblock {J1216+0709: A Radio Galaxy with Three Episodes of AGN Jet Activity}.
\newblock {\em \apj} {\bf 2016}, {\em 826},~132.
\newblock {\url{https://doi.org/10.3847/0004-637X/826/2/132}}.

\bibitem[{Chavan} et~al.(2023){Chavan}, {Dabhade}, and
  {Saikia}]{2023MNRAS.525L..87C}
{Chavan}, K.; {Dabhade}, P.; {Saikia}, D.J.
\newblock {A giant radio galaxy with three cycles of episodic jet activity from
  LoTSS DR2}.
\newblock {\em \mnras} {\bf 2023}, {\em 525},~L87--L92.
\newblock {\url{https://doi.org/10.1093/mnrasl/slad100}}.

\bibitem[{Massaro} et~al.(2009){Massaro}, {Giommi}, {Leto}, {Marchegiani},
  {Maselli}, {Perri}, {Piranomonte}, and {Sclavi}]{2009A&A...495..691M}
{Massaro}, E.; {Giommi}, P.; {Leto}, C.; {Marchegiani}, P.; {Maselli}, A.;
  {Perri}, M.; {Piranomonte}, S.; {Sclavi}, S.
\newblock {Roma-BZCAT: A multifrequency catalogue of blazars}.
\newblock {\em \aap} {\bf 2009}, {\em 495},~691--696.
\newblock {\url{https://doi.org/10.1051/0004-6361:200810161}}.

\bibitem[{Massaro} et~al.(2015){Massaro}, {Maselli}, {Leto}, {Marchegiani},
  {Perri}, {Giommi}, and {Piranomonte}]{2015Ap&SS.357...75M}
{Massaro}, E.; {Maselli}, A.; {Leto}, C.; {Marchegiani}, P.; {Perri}, M.;
  {Giommi}, P.; {Piranomonte}, S.
\newblock {The 5th edition of the Roma-BZCAT. A short presentation}.
\newblock {\em \apss} {\bf 2015}, {\em 357},~75.
\newblock {\url{https://doi.org/10.1007/s10509-015-2254-2}}.

\bibitem[{Adelman-McCarthy} et~al.(2008){Adelman-McCarthy}, {Ag{\"u}eros},
  {Allam}, {Allende Prieto}, {Anderson}, {Anderson}, {Annis}, {Bahcall},
  {Bailer-Jones}, {Baldry}, and et~al.]{2008ApJS..175..297A}
{Adelman-McCarthy}, J.K.; {Ag{\"u}eros}, M.A.; {Allam}, S.S.; {Allende Prieto},
  C.; {Anderson}, K.S.J.; {Anderson}, S.F.; {Annis}, J.; {Bahcall}, N.A.;
  {Bailer-Jones}, C.A.L.; {Baldry, A.A}.;  et~al.
\newblock {The Sixth Data Release of the Sloan Digital Sky Survey}.
\newblock {\em \apjs} {\bf 2008}, {\em 175},~297--313.
\newblock {\url{https://doi.org/10.1086/524984}}.


\bibitem[{Albareti} et~al.(2017){Albareti}, {Allende Prieto}, {Almeida},
  {Anders}, {Anderson}, {Andrews}, {Arag{\'o}n-Salamanca},
  {Argudo-Fern{\'a}ndez}, {Armengaud}, {Aubourg}, and
  et~al.]{2017ApJS..233...25A}
{Albareti}, F.D.; {Allende Prieto}, C.; {Almeida}, A.; {Anders}, F.;
  {Anderson}, S.; {Andrews}, B.H.; {Arag{\'o}n-Salamanca}, A.;
  {Argudo-Fern{\'a}ndez}, M.; {Armengaud}, E.; {Aubourg}, E.;  et~al.
\newblock {The 13th Data Release of the Sloan Digital Sky Survey: First
  Spectroscopic Data from the SDSS-IV Survey Mapping Nearby Galaxies at Apache
  Point Observatory}.
\newblock {\em \apjs} {\bf 2017}, {\em 233},~25.
\newblock {\url{https://doi.org/10.3847/1538-4365/aa8992}}.

\bibitem[{Becker} et~al.(1995){Becker}, {White}, and
  {Helfand}]{1995ApJ...450..559B}
{Becker}, R.H.; {White}, R.L.; {Helfand}, D.J.
\newblock {The FIRST Survey: Faint Images of the Radio Sky at Twenty
  Centimeters}.
\newblock {\em \apj} {\bf 1995}, {\em 450},~559.
\newblock {\url{https://doi.org/10.1086/176166}}.



\bibitem[{White} et~al.(1997){White}, {Becker}, {Helfand}, and
  {Gregg}]{1997ApJ...475..479W}
{White}, R.L.; {Becker}, R.H.; {Helfand}, D.J.; {Gregg}, M.D.
\newblock {A Catalog of 1.4 GHz Radio Sources from the FIRST Survey}.
\newblock {\em \apj} {\bf 1997}, {\em 475},~479--493.
\newblock {\url{https://doi.org/10.1086/303564}}.




\bibitem[{Fanaroff} and {Riley}(1974)]{1974MNRAS.167P..31F}
{Fanaroff}, B.L.; {Riley}, J.M.
\newblock {The morphology of extragalactic radio sources of high and low
  luminosity}.
\newblock {\em \mnras} {\bf 1974}, {\em 167},~31P--36P.
\newblock {\url{https://doi.org/10.1093/mnras/167.1.31P}}.

\bibitem[{Szabo} et~al.(2011){Szabo}, {Pierpaoli}, {Dong}, {Pipino}, and
  {Gunn}]{2011ApJ...736...21S}
{Szabo}, T.; {Pierpaoli}, E.; {Dong}, F.; {Pipino}, A.; {Gunn}, J.
\newblock {An Optical Catalog of Galaxy Clusters Obtained from an Adaptive
  Matched Filter Finder Applied to Sloan Digital Sky Survey Data Release 6}.
\newblock {\em \apj} {\bf 2011}, {\em 736},~21.
\newblock {\url{https://doi.org/10.1088/0004-637X/736/1/21}}.

\bibitem[{Mahatma}(2023)]{2023Galax..11...74M}
{Mahatma}, V.H.
\newblock {The Dynamics and Energetics of Remnant and Restarting RLAGN}.
\newblock {\em Galaxies} {\bf 2023}, {\em 11},~74.
\newblock {\url{https://doi.org/10.3390/galaxies11030074}}.

\bibitem[{Blandford} and {Znajek}(1977)]{1977MNRAS.179..433B}
{Blandford}, R.D.; {Znajek}, R.L.
\newblock {Electromagnetic extraction of energy from Kerr black holes.}
\newblock {\em \mnras} {\bf 1977}, {\em 179},~433--456.
\newblock {\url{https://doi.org/10.1093/mnras/179.3.433}}.

\bibitem[{Volonteri} et~al.(2005){Volonteri}, {Madau}, {Quataert}, and
  {Rees}]{2005ApJ...620...69V}
{Volonteri}, M.; {Madau}, P.; {Quataert}, E.; {Rees}, M.J.
\newblock {The Distribution and Cosmic Evolution of Massive Black Hole Spins}.
\newblock {\em \apj} {\bf 2005}, {\em 620},~69--77.
\newblock {\url{https://doi.org/10.1086/426858}}.

\bibitem[{Gergely} and {Biermann}(2009)]{2009ApJ...697.1621G}
{Gergely}, L.{\'A}.; {Biermann}, P.L.
\newblock {The Spin-Flip Phenomenon in Supermassive Black hole binary mergers}.
\newblock {\em \apj} {\bf 2009}, {\em 697},~1621--1633.
\newblock {\url{https://doi.org/10.1088/0004-637X/697/2/1621}}.

\bibitem[{Tchekhovskoy} et~al.(2011){Tchekhovskoy}, {Narayan}, and
  {McKinney}]{2011MNRAS.418L..79T}
{Tchekhovskoy}, A.; {Narayan}, R.; {McKinney}, J.C.
\newblock {Efficient generation of jets from magnetically arrested accretion on
  a rapidly spinning black hole}.
\newblock {\em \mnras} {\bf 2011}, {\em 418},~L79--L83.
\newblock {\url{https://doi.org/10.1111/j.1745-3933.2011.01147.x}}.

\bibitem[{van Velzen} and {Falcke}(2013)]{2013A&A...557L...7V}
{van Velzen}, S.; {Falcke}, H.
\newblock {The contribution of spin to jet-disk coupling in black holes}.
\newblock {\em \aap} {\bf 2013}, {\em 557},~L7.
\newblock {\url{https://doi.org/10.1051/0004-6361/201322127}}.

\bibitem[{McNamara} et~al.(2011){McNamara}, {Rohanizadegan}, and
  {Nulsen}]{2011ApJ...727...39M}
{McNamara}, B.R.; {Rohanizadegan}, M.; {Nulsen}, P.E.J.
\newblock {Are Radio Active Galactic Nuclei Powered by Accretion or Black Hole
  Spin?}
\newblock {\em \apj} {\bf 2011}, {\em 727},~39.
\newblock {\url{https://doi.org/10.1088/0004-637X/727/1/39}}.

\bibitem[{Dabhade} et~al.(2020){Dabhade}, {Mahato}, {Bagchi}, {Saikia},
  {Combes}, {Sankhyayan}, {R{\"o}ttgering}, {Ho}, {Gaikwad}, {Raychaudhury},
  {Vaidya}, and {Guiderdoni}]{2020A&A...642A.153D}
{Dabhade}, P.; {Mahato}, M.; {Bagchi}, J.; {Saikia}, D.J.; {Combes}, F.;
  {Sankhyayan}, S.; {R{\"o}ttgering}, H.J.A.; {Ho}, L.C.; {Gaikwad}, M.;
  {Raychaudhury}, S.;  et~al.
\newblock {Search and analysis of giant radio galaxies with associated nuclei
  (SAGAN). I. New sample and multi-wavelength studies}.
\newblock {\em \aap} {\bf 2020}, {\em 642},~A153.
\newblock {\url{https://doi.org/10.1051/0004-6361/202038344}}.

\bibitem[{Wright}(2006)]{2006PASP..118.1711W}
{Wright}, E.L.
\newblock {A Cosmology Calculator for the World Wide Web}.
\newblock {\em \pasp} {\bf 2006}, {\em 118},~1711--1715.
\newblock {\url{https://doi.org/10.1086/510102}}.

\bibitem[{Beasley} and {Conway}(1995)]{1995ASPC...82..327B}
{Beasley}, A.J.; {Conway}, J.E.
\newblock {VLBI Phase-Referencing}.
\newblock In Proceedings of the Very Long Baseline Interferometry and the VLBA, Socorro, NM, USA, 23--30 June 1993; {Zensus}, J.A., {Diamond}, P.J., {Napier}, P.J., Eds.; Astronomical Society of the Pacific Conference Series; Astronomical Society of the Pacific (ASP): San Francisco, CA, USA, 1995; Volume~82, p. 327.

\bibitem[{Charlot} et~al.(2020){Charlot}, {Jacobs}, {Gordon}, {Lambert}, {de
  Witt}, {B{\"o}hm}, {Fey}, {Heinkelmann}, {Skurikhina}, {Titov}, {Arias},
  {Bolotin}, {Bourda}, {Ma}, {Malkin}, {Nothnagel}, {Mayer}, {MacMillan},
  {Nilsson}, and {Gaume}]{2020A&A...644A.159C}
{Charlot}, P.; {Jacobs}, C.S.; {Gordon}, D.; {Lambert}, S.; {de Witt}, A.;
  {B{\"o}hm}, J.; {Fey}, A.L.; {Heinkelmann}, R.; {Skurikhina}, E.; {Titov},
  O.;  et~al.
\newblock {The third realization of the International Celestial Reference Frame
  by very long baseline interferometry}.
\newblock {\em \aap} {\bf 2020}, {\em 644},~A159.
\newblock {\url{https://doi.org/10.1051/0004-6361/202038368}}.

\bibitem[{Deller} et~al.(2011){Deller}, {Brisken}, {Phillips}, {Morgan},
  {Alef}, {Cappallo}, {Middelberg}, {Romney}, {Rottmann}, {Tingay}, and
  {Wayth}]{2011PASP..123..275D}
{Deller}, A.T.; {Brisken}, W.F.; {Phillips}, C.J.; {Morgan}, J.; {Alef}, W.;
  {Cappallo}, R.; {Middelberg}, E.; {Romney}, J.; {Rottmann}, H.; {Tingay},
  S.J.;  et~al.
\newblock {DiFX-2: A More Flexible, Efficient, Robust, and Powerful Software
  Correlator}.
\newblock {\em \pasp} {\bf 2011}, {\em 123},~275.
\newblock {\url{https://doi.org/10.1086/658907}}.

\bibitem[{Cao} et~al.(2014){Cao}, {Frey}, {Gurvits}, {Yang}, {Hong}, {Paragi},
  {Deller}, and {Ivezi{\'c}}]{2014A&A...563A.111C}
{Cao}, H.M.; {Frey}, S.; {Gurvits}, L.I.; {Yang}, J.; {Hong}, X.Y.; {Paragi},
  Z.; {Deller}, A.T.; {Ivezi{\'c}}, {\v{Z}}.
\newblock {VLBI observations of the radio quasar J2228+0110 at z = 5.95 and
  other field sources in multiple-phase-centre mode}.
\newblock {\em \aap} {\bf 2014}, {\em 563},~A111.
\newblock {\url{https://doi.org/10.1051/0004-6361/201323328}}.

\bibitem[{Keimpema} et~al.(2015){Keimpema}, {Kettenis}, {Pogrebenko},
  {Campbell}, {Cim{\'o}}, {Duev}, {Eldering}, {Kruithof}, {van Langevelde},
  {Marchal}, {Molera Calv{\'e}s}, {Ozdemir}, {Paragi}, {Pidopryhora},
  {Szomoru}, and {Yang}]{2015ExA....39..259K}
{Keimpema}, A.; {Kettenis}, M.M.; {Pogrebenko}, S.V.; {Campbell}, R.M.;
  {Cim{\'o}}, G.; {Duev}, D.A.; {Eldering}, B.; {Kruithof}, N.; {van
  Langevelde}, H.J.; {Marchal}, D.;  et~al.
\newblock {The SFXC software correlator for very long baseline interferometry:
  Algorithms and implementation}.
\newblock {\em Exp. Astron.} {\bf 2015}, {\em 39},~259--279.
\newblock {\url{https://doi.org/10.1007/s10686-015-9446-1}}.

\bibitem[{Greisen}(2003)]{2003ASSL..285..109G}
{Greisen}, E.W.
\newblock {AIPS, the VLA, and the VLBA}.
\newblock In\emph{ Information Handling in Astronomy---Historical
  Vistas}; {Heck}, A., Ed.; Astrophysics and Space Science
  Library; Springer: Berlin/Heidelberg, Germany, %We added the location of publisher. Please confirm - CONFIRMED
  2003; Volume 285,  p. 109.
\newblock {\url{https://doi.org/10.1007/0-306-48080-8_7}}.

\bibitem[{Diamond}(1995)]{1995ASPC...82..227D}
{Diamond}, P.J.
\newblock {VLBI Data Reduction in Practice}.
\newblock In Proceedings of the Very Long Baseline Interferometry and the VLBA, Socorro, NM, USA, 23--30 June 1993; {Zensus}, J.A., {Diamond}, P.J., {Napier}, P.J., Eds.; Astronomical Society of the Pacific Conference Series; Astronomical Society of the Pacific (ASP): San Francisco, CA, USA, 1995; Volume~82,  p. 227.

\bibitem[{Schwab} and {Cotton}(1983)]{1983AJ.....88..688S}
{Schwab}, F.R.; {Cotton}, W.D.
\newblock {Global fringe search techniques for VLBI}.
\newblock {\em \aj} {\bf 1983}, {\em 88},~688--694.
\newblock {\url{https://doi.org/10.1086/113360}}.

\bibitem[{Shepherd} et~al.(1994){Shepherd}, {Pearson}, and
  {Taylor}]{1994BAAS...26..987S}
{Shepherd}, M.C.; {Pearson}, T.J.; {Taylor}, G.B.
\newblock {DIFMAP: an interactive program for synthesis imaging.}
\newblock {\em Bull. Am. Astron. Soc.} {\bf 1994}, {\em
  26},~987--989.

\bibitem[{H{\"o}gbom}(1974)]{1974A&AS...15..417H}
{H{\"o}gbom}, J.A.
\newblock {Aperture Synthesis with a Non-Regular Distribution of Interferometer
  Baselines}.
\newblock {\em \aaps} {\bf 1974}, {\em 15},~417.

\bibitem[{Pearson} and {Readhead}(1984)]{1984ARA&A..22...97P}
{Pearson}, T.J.; {Readhead}, A.C.S.
\newblock {Image Formation by Self-Calibration in Radio Astronomy}.
\newblock {\em \araa} {\bf 1984}, {\em 22},~97--130.
\newblock {\url{https://doi.org/10.1146/annurev.aa.22.090184.000525}}.

\bibitem[{Pearson}(1995)]{1995ASPC...82..267P}
{Pearson}, T.J.
\newblock {Non-Imaging Data Analysis}.
\newblock In Proceedings of the Very Long Baseline Interferometry and the VLBA, Socorro, NM, USA, 23--30 June 1993; {Zensus}, J.A., {Diamond}, P.J., {Napier}, P.J., Eds.; Astronomical Society of the Pacific Conference Series; Astronomical Society of the Pacific (ASP): San Francisco, CA, USA, 1995; Volume~82, p. 267.

\bibitem[{Fomalont}(1999)]{1999ASPC..180..301F}
{Fomalont}, E.B.
\newblock {Image Analysis}.
\newblock In Proceedings of the {Synthesis Imaging in Radio Astronomy II}, Socorro, NM, USA, 17--23 June 1998;
  {Taylor}, G.B., {Carilli}, C.L., {Perley}, R.A., Eds.; Astronomical Society of the Pacific Conference Series; Astronomical Society of the Pacific (ASP): San Francisco, CA, USA, 1999; Volume 180,  p. 301.

\bibitem[{Lee} et~al.(2008){Lee}, {Lobanov}, {Krichbaum}, {Witzel}, {Zensus},
  {Bremer}, {Greve}, and {Grewing}]{2008AJ....136..159L}
{Lee}, S.S.; {Lobanov}, A.P.; {Krichbaum}, T.P.; {Witzel}, A.; {Zensus}, A.;
  {Bremer}, M.; {Greve}, A.; {Grewing}, M.
\newblock {A Global 86 GHz VLBI Survey of Compact Radio Sources}.
\newblock {\em \aj} {\bf 2008}, {\em 136},~159--180.
\newblock {\url{https://doi.org/10.1088/0004-6256/136/1/159}}.

\bibitem[{Condon} et~al.(1982){Condon}, {Condon}, {Gisler}, and
  {Puschell}]{1982ApJ...252..102C}
{Condon}, J.J.; {Condon}, M.A.; {Gisler}, G.; {Puschell}, J.J.
\newblock {Strong radio sources in bright spiral galaxies. II. Rapid star
  formation and galaxy-galaxy interactions.}
\newblock {\em \apj} {\bf 1982}, {\em 252},~102--124.
\newblock {\url{https://doi.org/10.1086/159538}}.

\bibitem[{Mart{\'\i}-Vidal} et~al.(2010){Mart{\'\i}-Vidal}, {Ros},
  {P{\'e}rez-Torres}, {Guirado}, {Jim{\'e}nez-Monferrer}, and
  {Marcaide}]{2010A&A...515A..53M}
{Mart{\'\i}-Vidal}, I.; {Ros}, E.; {P{\'e}rez-Torres}, M.A.; {Guirado}, J.C.;
  {Jim{\'e}nez-Monferrer}, S.; {Marcaide}, J.M.
\newblock {Coherence loss in phase-referenced VLBI observations}.
\newblock {\em \aap} {\bf 2010}, {\em 515},~A53.
\newblock {\url{https://doi.org/10.1051/0004-6361/201014203}}.

\bibitem[{Gab{\'a}nyi} et~al.(2019){Gab{\'a}nyi}, {Frey}, {Satyapal},
  {Constantin}, and {Pfeifle}]{2019A&A...630L...5G}
{Gab{\'a}nyi}, K.{\'E}.; {Frey}, S.; {Satyapal}, S.; {Constantin}, A.;
  {Pfeifle}, R.W.
\newblock {Very long baseline interferometry observation of the triple AGN
  candidate J0849+1114}.
\newblock {\em \aap} {\bf 2019}, {\em 630},~L5.
\newblock {\url{https://doi.org/10.1051/0004-6361/201936519}}.

\bibitem[{Krezinger} et~al.(2022){Krezinger}, {Perger}, {Gab{\'a}nyi}, {Frey},
  {Gurvits}, {Paragi}, {An}, {Zhang}, {Cao}, and
  {Sbarrato}]{2022ApJS..260...49K}
{Krezinger}, M.; {Perger}, K.; {Gab{\'a}nyi}, K.{\'E}.; {Frey}, S.; {Gurvits},
  L.I.; {Paragi}, Z.; {An}, T.; {Zhang}, Y.; {Cao}, H.; {Sbarrato}, T.
\newblock {Radio-loud Quasars above Redshift 4: Very Long Baseline
  Interferometry (VLBI) Imaging of an Extended Sample}.
\newblock {\em \apjs} {\bf 2022}, {\em 260},~49.
\newblock {\url{https://doi.org/10.3847/1538-4365/ac63b8}}.

\bibitem[{Krezinger} et~al.(2024){Krezinger}, {Baldini}, {Giroletti},
  {Sbarrato}, {Ghisellini}, {Giovannini}, {An}, {Gab{\'a}nyi}, and
  {Frey}]{2024A&A...690A.321K}
{Krezinger}, M.; {Baldini}, G.; {Giroletti}, M.; {Sbarrato}, T.; {Ghisellini},
  G.; {Giovannini}, G.; {An}, T.; {Gab{\'a}nyi}, K.{\'E}.; {Frey}, S.
\newblock {Revealing faint compact radio jets at redshifts above 5 with very
  long baseline interferometry}.
\newblock {\em \aap} {\bf 2024}, {\em 690},~A321.
\newblock {\url{https://doi.org/10.1051/0004-6361/202451025}}.

\bibitem[{Kewley} et~al.(2000){Kewley}, {Heisler}, {Dopita}, {Sutherland},
  {Norris}, {Reynolds}, and {Lumsden}]{2000ApJ...530..704K}
{Kewley}, L.J.; {Heisler}, C.A.; {Dopita}, M.A.; {Sutherland}, R.; {Norris},
  R.P.; {Reynolds}, J.; {Lumsden}, S.
\newblock {Compact Radio Emission from Warm Infrared Galaxies}.
\newblock {\em \apj} {\bf 2000}, {\em 530},~704--718.
\newblock {\url{https://doi.org/10.1086/308397}}.

\bibitem[{Krezinger} et~al.(2020){Krezinger}, {Frey}, {Paragi}, and
  {Deane}]{2020Symm...12..527K}
{Krezinger}, M.; {Frey}, S.; {Paragi}, Z.; {Deane}, R.
\newblock {High-Resolution Radio Observations of Five Optically Selected Type 2
  Quasars}.
\newblock {\em Symmetry} {\bf 2020}, {\em 12},~527.
\newblock {\url{https://doi.org/10.3390/sym12040527}}.

\bibitem[{Wang} et~al.(2023){Wang}, {An}, {Cheng}, {Ho}, {Kellermann}, {Baan},
  {Yang}, and {Zhang}]{2023MNRAS.518...39W}
{Wang}, A.; {An}, T.; {Cheng}, X.; {Ho}, L.C.; {Kellermann}, K.I.; {Baan},
  W.A.; {Yang}, J.; {Zhang}, Y.
\newblock {VLBI observations of a sample of Palomar-Green quasars - I.
  Parsec-scale morphology}.
\newblock {\em \mnras} {\bf 2023}, {\em 518},~39--53.
\newblock {\url{https://doi.org/10.1093/mnras/stac3091}}.

\bibitem[{Gaia Collaboration} et~al.(2016){Gaia Collaboration}, {Prusti}, {de
  Bruijne}, {Brown}, {Vallenari}, {Babusiaux}, {Bailer-Jones}, {Bastian},
  {Biermann}, {Evans}, {Eyer}, {Jansen}, {Jordi}, {Klioner}, {Lammers},
  {Lindegren}, {Luri}, {Mignard}, {Milligan}, {Panem}, {Poinsignon},
  {Pourbaix}, {Randich}, {Sarri}, {Sartoretti}, {Siddiqui}, {Soubiran},
  {Valette}, {van Leeuwen}, {Walton}, {Aerts}, {Arenou}, {Cropper}, {Drimmel},
  {H{\o}g}, {Katz}, {Lattanzi}, {O'Mullane}, {Grebel}, {Holland}, {Huc},
  {Passot}, {Bramante}, {Cacciari}, {Casta{\~n}eda}, {Chaoul}, {Cheek}, {De
  Angeli}, {Fabricius}, {Guerra}, {Hern{\'a}ndez}, {Jean-Antoine-Piccolo},
  {Masana}, {Messineo}, {Mowlavi}, {Nienartowicz}, {Ord{\'o}{\~n}ez-Blanco},
  {Panuzzo}, {Portell}, {Richards}, {Riello}, {Seabroke}, {Tanga},
  {Th{\'e}venin}, {Torra}, {Els}, {Gracia-Abril}, {Comoretto},
  {Garcia-Reinaldos}, {Lock}, {Mercier}, {Altmann}, {Andrae}, {Astraatmadja},
  {Bellas-Velidis}, {Benson}, {Berthier}, {Blomme}, {Busso}, {Carry},
  {Cellino}, {Clementini}, {Cowell}, {Creevey}, {Cuypers}, {Davidson}, {De
  Ridder}, {de Torres}, {Delchambre}, {Dell'Oro}, {Ducourant}, {Fr{\'e}mat},
  {Garc{\'\i}a-Torres}, {Gosset}, {Halbwachs}, {Hambly}, {Harrison}, {Hauser},
  {Hestroffer}, {Hodgkin}, {Huckle}, {Hutton}, {Jasniewicz}, {Jordan},
  {Kontizas}, {Korn}, {Lanzafame}, {Manteiga}, {Moitinho}, {Muinonen},
  {Osinde}, {Pancino}, {Pauwels}, {Petit}, {Recio-Blanco}, {Robin}, {Sarro},
  {Siopis}, {Smith}, {Smith}, {Sozzetti}, {Thuillot}, {van Reeven}, {Viala},
  {Abbas}, {Abreu Aramburu}, {Accart}, {Aguado}, {Allan}, {Allasia},
  {Altavilla}, {{\'A}lvarez}, {Alves}, {Anderson}, {Andrei}, {Anglada Varela},
  {Antiche}, {Antoja}, {Ant{\'o}n}, {Arcay}, {Atzei}, {Ayache}, {Bach},
  {Baker}, {Balaguer-N{\'u}{\~n}ez}, {Barache}, {Barata}, {Barbier}, {Barblan},
  {Baroni}, {Barrado y Navascu{\'e}s}, {Barros}, {Barstow}, {Becciani},
  {Bellazzini}, {Bellei}, {Bello Garc{\'\i}a}, {Belokurov}, {Bendjoya},
  {Berihuete}, {Bianchi}, {Bienaym{\'e}}, {Billebaud}, {Blagorodnova},
  {Blanco-Cuaresma}, {Boch}, {Bombrun}, {Borrachero}, {Bouquillon}, {Bourda},
  {Bouy}, {Bragaglia}, {Breddels}, {Brouillet}, {Br{\"u}semeister},
  {Bucciarelli}, {Budnik}, {Burgess}, {Burgon}, {Burlacu}, {Busonero}, {Buzzi},
  {Caffau}, {Cambras}, {Campbell}, {Cancelliere}, {Cantat-Gaudin}, {Carlucci},
  {Carrasco}, {Castellani}, {Charlot}, {Charnas}, {Charvet}, {Chassat},
  {Chiavassa}, {Clotet}, {Cocozza}, {Collins}, {Collins}, {Costigan}, {Crifo},
  {Cross}, {Crosta}, {Crowley}, {Dafonte}, {Damerdji}, {Dapergolas}, {David},
  {David}, {De Cat}, {de Felice}, {de Laverny}, {De Luise}, {De March}, {de
  Martino}, {de Souza}, {Debosscher}, {del Pozo}, {Delbo}, {Delgado},
  {Delgado}, {di Marco}, {Di Matteo}, {Diakite}, {Distefano}, {Dolding}, {Dos
  Anjos}, {Drazinos}, {Dur{\'a}n}, {Dzigan}, {Ecale}, {Edvardsson}, {Enke},
  {Erdmann}, {Escolar}, {Espina}, {Evans}, {Eynard Bontemps}, {Fabre},
  {Fabrizio}, {Faigler}, {Falc{\~a}o}, {Farr{\`a}s Casas}, {Faye}, {Federici},
  {Fedorets}, {Fern{\'a}ndez-Hern{\'a}ndez}, {Fernique}, {Fienga}, {Figueras},
  {Filippi}, {Findeisen}, {Fonti}, {Fouesneau}, {Fraile}, {Fraser}, {Fuchs},
  {Furnell}, {Gai}, {Galleti}, {Galluccio}, {Garabato}, {Garc{\'\i}a-Sedano},
  {Gar{\'e}}, {Garofalo}, {Garralda}, {Gavras}, {Gerssen}, {Geyer}, {Gilmore},
  {Girona}, {Giuffrida}, {Gomes}, {Gonz{\'a}lez-Marcos},
  {Gonz{\'a}lez-N{\'u}{\~n}ez}, {Gonz{\'a}lez-Vidal}, {Granvik}, {Guerrier},
  {Guillout}, {Guiraud}, {G{\'u}rpide}, {Guti{\'e}rrez-S{\'a}nchez}, {Guy},
  {Haigron}, {Hatzidimitriou}, {Haywood}, {Heiter}, {Helmi}, {Hobbs},
  {Hofmann}, {Holl}, {Holland}, {Hunt}, {Hypki}, {Icardi}, {Irwin}, {Jevardat
  de Fombelle}, {Jofr{\'e}}, {Jonker}, {Jorissen}, {Julbe}, {Karampelas},
  {Kochoska}, {Kohley}, {Kolenberg}, {Kontizas}, {Koposov}, {Kordopatis},
  {Koubsky}, {Kowalczyk}, {Krone-Martins}, {Kudryashova}, {Kull}, {Bachchan},
  {Lacoste-Seris}, {Lanza}, {Lavigne}, {Le Poncin-Lafitte}, {Lebreton},
  {Lebzelter}, {Leccia}, {Leclerc}, {Lecoeur-Taibi}, {Lemaitre}, {Lenhardt},
  {Leroux}, {Liao}, {Licata}, {Lindstr{\o}m}, {Lister}, {Livanou}, {Lobel},
  {L{\"o}ffler}, {L{\'o}pez}, {Lopez-Lozano}, {Lorenz}, {Loureiro},
  {MacDonald}, {Magalh{\~a}es Fernandes}, {Managau}, {Mann}, {Mantelet},
  {Marchal}, {Marchant}, {Marconi}, {Marie}, {Marinoni}, {Marrese},
  {Marschalk{\'o}}, {Marshall}, {Mart{\'\i}n-Fleitas}, {Martino}, {Mary},
  {Matijevi{\v{c}}}, {Mazeh}, {McMillan}, {Messina}, {Mestre}, {Michalik},
  {Millar}, {Miranda}, {Molina}, {Molinaro}, {Molinaro}, {Moln{\'a}r},
  {Moniez}, {Montegriffo}, {Monteiro}, {Mor}, {Mora}, {Morbidelli}, {Morel},
  {Morgenthaler}, {Morley}, {Morris}, {Mulone}, {Muraveva}, {Musella},
  {Narbonne}, {Nelemans}, {Nicastro}, {Noval}, {Ord{\'e}novic},
  {Ordieres-Mer{\'e}}, {Osborne}, {Pagani}, {Pagano}, {Pailler}, {Palacin},
  {Palaversa}, {Parsons}, {Paulsen}, {Pecoraro}, {Pedrosa}, {Pentik{\"a}inen},
  {Pereira}, {Pichon}, {Piersimoni}, {Pineau}, {Plachy}, {Plum}, {Poujoulet},
  {Pr{\v{s}}a}, {Pulone}, {Ragaini}, {Rago}, {Rambaux}, {Ramos-Lerate},
  {Ranalli}, {Rauw}, {Read}, {Regibo}, {Renk}, {Reyl{\'e}}, {Ribeiro},
  {Rimoldini}, {Ripepi}, {Riva}, {Rixon}, {Roelens}, {Romero-G{\'o}mez},
  {Rowell}, {Royer}, {Rudolph}, {Ruiz-Dern}, {Sadowski}, {Sagrist{\`a}
  Sell{\'e}s}, {Sahlmann}, {Salgado}, {Salguero}, {Sarasso}, {Savietto},
  {Schnorhk}, {Schultheis}, {Sciacca}, {Segol}, {Segovia}, {Segransan},
  {Serpell}, {Shih}, {Smareglia}, {Smart}, {Smith}, {Solano}, {Solitro},
  {Sordo}, {Soria Nieto}, {Souchay}, {Spagna}, {Spoto}, {Stampa}, {Steele},
  {Steidelm{\"u}ller}, {Stephenson}, {Stoev}, {Suess}, {S{\"u}veges}, {Surdej},
  {Szabados}, {Szegedi-Elek}, {Tapiador}, {Taris}, {Tauran}, {Taylor},
  {Teixeira}, {Terrett}, {Tingley}, {Trager}, {Turon}, {Ulla}, {Utrilla},
  {Valentini}, {van Elteren}, {Van Hemelryck}, {van Leeuwen}, {Varadi},
  {Vecchiato}, {Veljanoski}, {Via}, {Vicente}, {Vogt}, {Voss}, {Votruba},
  {Voutsinas}, {Walmsley}, {Weiler}, {Weingrill}, {Werner}, {Wevers},
  {Whitehead}, {Wyrzykowski}, {Yoldas}, {{\v{Z}}erjal}, {Zucker}, {Zurbach},
  {Zwitter}, {Alecu}, {Allen}, {Allende Prieto}, {Amorim},
  {Anglada-Escud{\'e}}, {Arsenijevic}, {Azaz}, {Balm}, {Beck}, {Bernstein},
  {Bigot}, {Bijaoui}, {Blasco}, {Bonfigli}, {Bono}, {Boudreault}, {Bressan},
  {Brown}, {Brunet}, {Bunclark}, {Buonanno}, {Butkevich}, {Carret}, {Carrion},
  {Chemin}, {Ch{\'e}reau}, {Corcione}, {Darmigny}, {de Boer}, {de Teodoro}, {de
  Zeeuw}, {Delle Luche}, {Domingues}, {Dubath}, {Fodor}, {Fr{\'e}zouls},
  {Fries}, {Fustes}, {Fyfe}, {Gallardo}, {Gallegos}, {Gardiol}, {Gebran},
  {Gomboc}, {G{\'o}mez}, {Grux}, {Gueguen}, {Heyrovsky}, {Hoar}, {Iannicola},
  {Isasi Parache}, {Janotto}, {Joliet}, {Jonckheere}, {Keil}, {Kim},
  {Klagyivik}, {Klar}, {Knude}, {Kochukhov}, {Kolka}, {Kos}, {Kutka}, {Lainey},
  {LeBouquin}, {Liu}, {Loreggia}, {Makarov}, {Marseille}, {Martayan},
  {Martinez-Rubi}, {Massart}, {Meynadier}, {Mignot}, {Munari}, {Nguyen},
  {Nordlander}, {Ocvirk}, {O'Flaherty}, {Olias Sanz}, {Ortiz}, {Osorio},
  {Oszkiewicz}, {Ouzounis}, {Palmer}, {Park}, {Pasquato}, {Peltzer}, {Peralta},
  {P{\'e}turaud}, {Pieniluoma}, {Pigozzi}, {Poels}, {Prat}, {Prod'homme},
  {Raison}, {Rebordao}, {Risquez}, {Rocca-Volmerange}, {Rosen}, {Ruiz-Fuertes},
  {Russo}, {Sembay}, {Serraller Vizcaino}, {Short}, {Siebert}, {Silva},
  {Sinachopoulos}, {Slezak}, {Soffel}, {Sosnowska}, {Strai{\v{z}}ys}, {ter
  Linden}, {Terrell}, {Theil}, {Tiede}, {Troisi}, {Tsalmantza}, {Tur},
  {Vaccari}, {Vachier}, {Valles}, {Van Hamme}, {Veltz}, {Virtanen}, {Wallut},
  {Wichmann}, {Wilkinson}, {Ziaeepour}, and {Zschocke}]{2016A&A...595A...1G}
{Gaia Collaboration}.; {Prusti}, T.; {de Bruijne}, J.H.J.; {Brown}, A.G.A.;
  {Vallenari}, A.; {Babusiaux}, C.; {Bailer-Jones}, C.A.L.; {Bastian}, U.;
  {Biermann}, M.; {Evans}, D.W.;  et~al.
\newblock {The Gaia mission}.
\newblock {\em \aap} {\bf 2016}, {\em 595},~A1.
\newblock {\url{https://doi.org/10.1051/0004-6361/201629272}}.

\bibitem[{Gaia Collaboration} et~al.(2023){Gaia Collaboration}, {Vallenari},
  {Brown}, {Prusti}, {de Bruijne}, {Arenou}, {Babusiaux}, {Biermann},
  {Creevey}, {Ducourant}, and et~al.]{2023A&A...674A...1G}
{Gaia Collaboration}.; {Vallenari}, A.; {Brown}, A.G.A.; {Prusti}, T.; {de
  Bruijne}, J.H.J.; {Arenou}, F.; {Babusiaux}, C.; {Biermann}, M.; {Creevey},
  O.L.; {Ducourant}, C.;  et~al.
\newblock {Gaia Data Release 3. Summary of the content and survey properties}.
\newblock {\em \aap} {\bf 2023}, {\em 674},~A1.
\newblock {\url{https://doi.org/10.1051/0004-6361/202243940}}.

\bibitem[{Orosz} and {Frey}(2013)]{2013A&A...553A..13O}
{Orosz}, G.; {Frey}, S.
\newblock {Optical-radio positional offsets for active galactic nuclei}.
\newblock {\em \aap} {\bf 2013}, {\em 553},~A13.
\newblock {\url{https://doi.org/10.1051/0004-6361/201321279}}.

\bibitem[{Marecki} and {Swoboda}(2011)]{2011A&A...525A...6M}
{Marecki}, A.; {Swoboda}, B.
\newblock {The transition from quasar radio-loud to radio-quiet state in the
  framework of the black hole scalability hypothesis}.
\newblock {\em \aap} {\bf 2011}, {\em 525},~A6.
\newblock {\url{https://doi.org/10.1051/0004-6361/201015461}}.

\bibitem[{Wright} et~al.(2010){Wright}, {Eisenhardt}, {Mainzer}, {Ressler},
  {Cutri}, {Jarrett}, {Kirkpatrick}, {Padgett}, {McMillan}, {Skrutskie},
  {Stanford}, {Cohen}, {Walker}, {Mather}, {Leisawitz}, {Gautier}, {McLean},
  {Benford}, {Lonsdale}, {Blain}, {Mendez}, {Irace}, {Duval}, {Liu}, {Royer},
  {Heinrichsen}, {Howard}, {Shannon}, {Kendall}, {Walsh}, {Larsen}, {Cardon},
  {Schick}, {Schwalm}, {Abid}, {Fabinsky}, {Naes}, and
  {Tsai}]{2010AJ....140.1868W}
{Wright}, E.L.; {Eisenhardt}, P.R.M.; {Mainzer}, A.K.; {Ressler}, M.E.;
  {Cutri}, R.M.; {Jarrett}, T.; {Kirkpatrick}, J.D.; {Padgett}, D.; {McMillan},
  R.S.; {Skrutskie}, M.;  et~al.
\newblock {The Wide-field Infrared Survey Explorer (WISE): Mission Description
  and Initial On-orbit Performance}.
\newblock {\em \aj} {\bf 2010}, {\em 140},~1868--1881.
\newblock {\url{https://doi.org/10.1088/0004-6256/140/6/1868}}.

\bibitem[{Cutri} et~al.(2021){Cutri}, {Wright}, {Conrow}, {Fowler},
  {Eisenhardt}, {Grillmair}, {Kirkpatrick}, {Masci}, {McCallon}, {Wheelock},
  {Fajardo-Acosta}, {Yan}, {Benford}, {Harbut}, {Jarrett}, {Lake}, {Leisawitz},
  {Ressler}, {Stanford}, {Tsai}, {Liu}, {Helou}, {Mainzer}, {Gettngs},
  {Gonzalez}, {Hoffman}, {Marsh}, {Padgett}, {Skrutskie}, {Beck}, {Papin}, and
  {Wittman}]{2014yCat.2328....0C}
{Cutri}, R.M.; {Wright}, E.L.; {Conrow}, T.; {Fowler}, J.W.; {Eisenhardt},
  P.R.M.; {Grillmair}, C.; {Kirkpatrick}, J.D.; {Masci}, F.; {McCallon}, H.L.;
  {Wheelock}, S.L.;  et~al.
\newblock {VizieR Online Data Catalog: AllWISE Data Release II/328, \url{https://cdsarc.cds.unistra.fr/viz-bin/cat/II/328}, accessed on 14 December 2024,} {\bf 2023}.

\bibitem[{Marecki}(2012)]{2012A&A...544L...2M}
{Marecki}, A.
\newblock {Activity restart - a key to explaining the morphology of J1211+743}.
\newblock {\em \aap} {\bf 2012}, {\em 544},~L2.
\newblock {\url{https://doi.org/10.1051/0004-6361/201219638}}.

\bibitem[{Readhead}(1994)]{1994ApJ...426...51R}
{Readhead}, A.C.S.
\newblock {Equipartition Brightness Temperature and the Inverse Compton
  Catastrophe}.
\newblock {\em \apj} {\bf 1994}, {\em 426},~51.
\newblock {\url{https://doi.org/10.1086/174038}}.

\bibitem[{Homan} et~al.(2021){Homan}, {Cohen}, {Hovatta}, {Kellermann},
  {Kovalev}, {Lister}, {Popkov}, {Pushkarev}, {Ros}, and
  {Savolainen}]{2021ApJ...923...67H}
{Homan}, D.C.; {Cohen}, M.H.; {Hovatta}, T.; {Kellermann}, K.I.; {Kovalev},
  Y.Y.; {Lister}, M.L.; {Popkov}, A.V.; {Pushkarev}, A.B.; {Ros}, E.;
  {Savolainen}, T.
\newblock {MOJAVE. XIX. Brightness Temperatures and Intrinsic Properties of
  Blazar Jets}.
\newblock {\em \apj} {\bf 2021}, {\em 923},~67.
\newblock {\url{https://doi.org/10.3847/1538-4357/ac27af}}.

\bibitem[{Koz{\'a}k} et~al.(2024){Koz{\'a}k}, {Frey}, and
  {Gab{\'a}nyi}]{2024Galax..12....8K}
{Koz{\'a}k}, B.; {Frey}, S.; {Gab{\'a}nyi}, K.{\'E}.
\newblock {Superluminal Motion and Jet Parameters in the Gamma-ray-Emitting
  Narrow-Line Seyfert 1 Galaxy TXS 1206+549}.
\newblock {\em Galaxies} {\bf 2024}, {\em 12},~8.
\newblock {\url{https://doi.org/10.3390/galaxies12010008}}.

\bibitem[{Xie} et~al.(2024){Xie}, {Ba{\~n}ados}, {Belladitta}, {Mazzucchelli},
  {Schindler}, {Davies}, and {Venemans}]{2024ApJ...964...98X}
{Xie}, Z.L.; {Ba{\~n}ados}, E.; {Belladitta}, S.; {Mazzucchelli}, C.;
  {Schindler}, J.T.; {Davies}, F.; {Venemans}, B.P.
\newblock {Recognizing Blazars Using Radio Morphology from the VLA Sky Survey}.
\newblock {\em \apj} {\bf 2024}, {\em 964},~98.
\newblock {\url{https://doi.org/10.3847/1538-4357/ad20d3}}.

\bibitem[{Mateos} et~al.(2012){Mateos}, {Alonso-Herrero}, {Carrera}, {Blain},
  {Watson}, {Barcons}, {Braito}, {Severgnini}, {Donley}, and
  {Stern}]{2012MNRAS.426.3271M}
{Mateos}, S.; {Alonso-Herrero}, A.; {Carrera}, F.J.; {Blain}, A.; {Watson},
  M.G.; {Barcons}, X.; {Braito}, V.; {Severgnini}, P.; {Donley}, J.L.; {Stern},
  D.
\newblock {Using the Bright Ultrahard XMM-Newton survey to define an IR
  selection of luminous AGN based on WISE colours}.
\newblock {\em \mnras} {\bf 2012}, {\em 426},~3271--3281.
\newblock {\url{https://doi.org/10.1111/j.1365-2966.2012.21843.x}}.

\bibitem[{Massaro} et~al.(2012){Massaro}, {D'Abrusco}, {Tosti}, {Ajello},
  {Gasparrini}, {Grindlay}, and {Smith}]{2012ApJ...750..138M}
{Massaro}, F.; {D'Abrusco}, R.; {Tosti}, G.; {Ajello}, M.; {Gasparrini}, D.;
  {Grindlay}, J.E.; {Smith}, H.A.
\newblock {The WISE Gamma-Ray Strip Parameterization: The Nature of the
  Gamma-Ray Active Galactic Nuclei of Uncertain Type}.
\newblock {\em \apj} {\bf 2012}, {\em 750},~138.
\newblock {\url{https://doi.org/10.1088/0004-637X/750/2/138}}.

\bibitem[{Lindegren} et~al.(2012){Lindegren}, {Lammers}, {Hobbs}, {O'Mullane},
  {Bastian}, and {Hern{\'a}ndez}]{2012A&A...538A..78L}
{Lindegren}, L.; {Lammers}, U.; {Hobbs}, D.; {O'Mullane}, W.; {Bastian}, U.;
  {Hern{\'a}ndez}, J.
\newblock {The astrometric core solution for the Gaia mission. Overview of
  models, algorithms, and software implementation}.
\newblock {\em \aap} {\bf 2012}, {\em 538},~A78.
\newblock {\url{https://doi.org/10.1051/0004-6361/201117905}}.

\end{thebibliography}

\PublishersNote{}
\end{adjustwidth}
\end{document}